\documentclass[
  aps,
  prl,
  reprint,
  noeprint,
  longbibliography,
  floatfix,
  superscriptaddress,
  nofootinbib
]{revtex4-2}
\usepackage[utf8]{inputenc}
\usepackage[T1]{fontenc}
\usepackage{amsmath}
\usepackage{amssymb}
\usepackage{physics}
\usepackage{mathtools}
\usepackage{array}
\usepackage[dvipsnames]{xcolor}
    \definecolor{darkgreen}{rgb}{0,0.5,0}
    \definecolor{darkblue}{rgb}{0,0,0.6}
    \definecolor{purple}{rgb}{0.4,.2,0.7}
	\definecolor{linkblue}{rgb}{0.0,0.0,0.6}
	\definecolor{citepurple}{rgb}{0.35,0.2,0.55}
\usepackage[hyperfootnotes = false, 
	colorlinks = true,
	linkcolor = darkblue, 
	citecolor = purple, 
	urlcolor = linkblue]{hyperref}
\usepackage{relsize}
\usepackage{CJKutf8}
\usepackage{balance}

\newcommand{\ii}{\mathrm{i}}
\newcommand{\ee}{e}

\newcommand{\KL}{\mathsf{K}} 
\newcommand{\sJ}{\mathcal{J}} 

\begin{document}

\title{
    From integrability to many-body quantum chaos through a Markovian bath
}

\author{Xianlong Liu ({\begin{CJK}{UTF8}{gbsn}刘显龙\end{CJK}})}
\email{xianlong\_liu@sjtu.edu.cn}
\affiliation{Shanghai Center for Complex Physics, School of Physics and Astronomy,
    Shanghai Jiao Tong University, Shanghai 200240, China
}

\author{Antonio M. Garc\'ia-Garc\'ia}
\email{amgg@sjtu.edu.cn}
\affiliation{Shanghai Center for Complex Physics, School of Physics and Astronomy,
	Shanghai Jiao Tong University, Shanghai 200240, China
}

\begin{abstract}
Recent examples have confirmed the common belief that quantum chaos is always suppressed in the presence of an environment. Here we show that this is not always the case. We compute the Lyapunov exponent of a $q$-body Majorana Sachdev-Ye-Kitaev (SYK) model coupled to a Markovian bath. Provided that the jump operators describing the bath are isotropic random $k > 2$-body Majoranas fields, the Lyapunov exponent is positive for any strength of the coupling to the bath. Interestingly, this also applies to $q=2$ where the SYK is integrable which indicates that many-body quantum chaos can be induced by the environment. Moreover, for $q > 2$, where the unitary dynamics is quantum chaotic, and sufficiently large $k$, the Lyapunov exponent increases with the coupling to the bath. Explicit analytical results are obtained in the $q=2$ and large $q$ limits. Our results put forward an alternate route to induce and control the generation of scrambling in quantum many-body systems which is of potential relevance in the design of quantum information devices.
\end{abstract}

\maketitle
One of the main motivations to study the dynamics of quantum chaotic systems is its universality, namely, the fact that vastly different systems, single-particle or many-body, share common dynamical features. One of them is the exponential growth of quantum uncertainty around the scrambling time \cite{larkin1969}, a short time scale, at a rate given by the Lyapunov exponent which for systems with a classical limit coincides with the classical Lyapunov exponent. The explicit analytical calculation  of this exponential growth in simple systems such as a particle in a random potential \cite{larkin1969} or kicked rotors \cite{berman1978} were fundamental building blocks for the establishment of a theory of quantum chaos. Another key result is the so called Bohigas-Giannoni-Schmmit’s conjecture \cite{bohigas1984} stating that for long times of the order of the Heisenberg time, the spectral correlations of quantum chaotic system are given by those of a random matrix so the dynamics becomes universal again in this region. This latter characterization is easier to implement given that the spectrum is straightforward to compute numerically and the exponential growth of the uncertainty, captured by the so-called out-of-time ordered correlation (OTOC) functions, is hard to compute, either analytically or numerically. 

However, this has changed more recently because of two seminal results: that the Lyapunov exponent is subject to a universal bound \cite{maldacena2015} which is saturated in field theories with a gravity dual and the introduction of the Sachdev-Ye-Kitaev (SYK) model \cite{kitaev2015,french1970,bohigas1971,maldacena2016,sachdev1993,benet2001}, a toy model of holography, and also for many-body quantum chaos \cite{kitaev2015,garcia2016}, consisting of $N$ Majoranas with random infinite range interactions in zero spatial dimension. The analytical tractability of the SYK model \cite{kitaev2015} has made possible not only to compute the Lyapunov exponent in many situations, both analytically \cite{maldacena2016,gu2022} and numerically \cite{kobrin2020,garcia2024d}, but also to investigate the evolution of OTOC at all times
scales \cite{garcia2026}. 

In recent years, there has been a growing interest in the study of the dynamics of quantum chaotic systems in contact with an environment \cite{zanardi2021,tuziemski2019,yoshida2019,syzranov2018,bergamasco2023,pengfei2023,altman2023} so the evolution is not in general unitary. The SYK model is also quite helpful in this case \cite{kulkarni2022,sa2022,garcia2023}. It has been shown \cite{garcia2024} that for the case of a Markovian environment, described by the Lindblad formalism \cite{lindblad1976}, quantum chaos is robust to the presence of dissipation but the Lyapunov exponent is gradually reduced until it vanishes at a certain coupling to the bath. If the dynamics of the SYK is not
chaotic, the Lyapunov exponent is always negative \cite{garcia2025} and the  environment only makes the decay faster so it always suppresses quantum chaos. Note that these results were obtained for the specific case of a Markovian bath characterized by a jump operator with a non random coupling and linear in the Majorana fields. For a non-Markovian environment resulting for instance from the coupling of an SYK to reservoir, given by another SYK, the Lyapunov exponent always decreases \cite{chen2017a} with the strength of the bath coupling though whether it vanishes or not depends on the details of the system-bath interaction. Finally, the combination of Markovian and
non-Markovian baths \cite{liu2026,pengfei2026} leads to an enhancement of the Lyapunov in certain
circumstances but always below the one corresponding to the Hermitian limit.
Transitions characterized by the vanishing of the Lyapunov exponent induced by increasing the coupling to the bath have also been reported \cite{altman2023,pengfei2023} beyond the SYK model. Therefore, in view of these results, it is natural to expect that the coupling to a bath always suppresses scrambling and reduces the Lyapunov exponent. 

The main aim of this paper is to provide conclusive evidence that this is not always the case. We show that, if the jump operators that define the Markovian bath are given by the product of a sufficiently large number $k$ of Majoranas with random coefficients, the Lyapunov exponent will increase with the coupling, with a maximum enhancement for $q \sim k$ in the case $q=4$, and will be positive even in the $q=2$ limit where the SYK model is non-quantum chaotic. We start our analysis by defining the model that we term Lindblad SYK.

\textit{Model and Schwinger-Dyson equations.--} The dynamics of a quantum system coupled to a Markovian bath can be described by the Lindblad equation
\begin{equation} \label{eq:Lindblad_eq}
    \dv{\rho}{t} = - \ii [H, \rho] + \sum_{a} \Bigl( 2 L_a \rho L_a^{\dagger} - \{ L_a^\dagger L_a, \rho \} \Bigr) ,
\end{equation}
where $\rho$ is the density matrix of the system. Let $\chi_{i}$ ($i = 1, \dots, N$) denote Majorana fermions obeying anti-commutation relation $\{\chi_{i}, \chi_{j} \} = \delta_{ij}$, the SYK Hamiltonian is given by \cite{kitaev2015}
\begin{equation}
H = \ii^{q(q-1)/2} \sum_{1 \leq j_1 < \dots j_q \leq N} J_{j_1 \dots j_q} \chi_{j_1} \dots \chi_{j_q} ,
\end{equation}
where the random couplings $J_{j_1 \dots j_q}$ are drawn from a Gaussian distribution with zero mean and variance
\begin{equation}
\Bigl\langle \bigl(J_{j_1 \dots j_q}\bigr)^2 \Bigr\rangle = \frac{(q-1)! J^2}{N^{q-1}} .
\end{equation}
We consider a Markovian bath characterized by $k$-body random quantum jump operators \cite{sa2022,kulkarni2022}
\begin{equation}
\label{eq:jump_opeartors_def}
L_a = \ii^{k(k-1)/2} \sum_{1 \leq j_1 < \dots < j_k \leq N} \ell_{j_1 \dots j_k}^{a} \chi_{j_1} \dots \chi_{j_k} ,
\end{equation}
where $a = 1,\dots, M$. The random couplings $\ell_{j_1 \dots j_k}^{a}$ are random numbers drawn from Gaussian distribution with zero mean and variance \cite{sa2022,kulkarni2022}
\begin{equation}
    \Bigl\langle \bigl(\ell_{j_1 \dots j_k}^{a}\bigr)^2 \Bigr\rangle = \frac{k! \gamma^2}{(2k)^{1/2} N^k} .
\end{equation}
We study this model in the large $N$ limit while keeping $m \equiv M/N$ fixed. Let $z \equiv t + \ii \tau$ denote the time argument on the Schwinger-Keldysh (SK) contour $\mathcal{C}$. We define the bi-local collective field
\begin{equation}
    G(z_1, z_2) \equiv - \frac{\ii}{N} \sum_{j=1}^{N} \chi_j(z_1) \chi_j(z_2) .
\end{equation}
The SK contour consists of the forward~($+$) and backward~($-$) branches, and different combinations give different kinds of the Green's functions. For instance, we have the greater Green's function $G^>(t, t')\equiv G_{-+}(t, t')$ and lesser Green's function $G^<(t, t') \equiv G_{+-}(t, t')$. Performing disorder averages on the random couplings and integrating out the Majorana fermions, we obtain the path integral
$
Z = \int \mathcal{D} G \mathcal{D} \Sigma \ee^{\ii S[G, \Sigma]} 
$,
with the action
\begin{align}
    \frac{2}{N} \ii S = 
    &\,
    \Tr_{\mathcal{C}} \log (\ii \partial - \Sigma) 
    - \int_{\mathcal{C}} \Sigma(z, z') G(z, z') \dd{z} \dd{z'} \nonumber \\
    &\,
    - \frac{\ii^q J^2}{q} \int_{\mathcal{C}} [G(z, z')]^{q} \dd{z} \dd{z'} 
    \nonumber \\
    &\, - 2 (-1)^k m \Tr_{\mathcal{C}} \log(1 + \frac{\ii^k \gamma^2}{(2k)^{1/2}} B) .
    \label{eq:action_full}
\end{align}
Here $\KL$ denotes the Lindblad kernel on the SK contour and $B$ is defined by a convolution,
\begin{align} 
    \label{eq:K_lindblad_kernel_one_fold}
    \KL(z_1, z_2) & \equiv \delta(t_1 - t_2) \begin{pmatrix}
    1 & 0 \\
    2 & 1
    \end{pmatrix} , 
    \\
    B(z_1, z_2) & \equiv \int_{\mathcal{C}} \KL(z_1, z) [G(z, z_2)]^{k} \dd{z} .
\end{align}
The Schwinger-Dyson (SD) equations are given by
\begin{equation}
    (\ii \partial - \Sigma) \circ G = 1_{\mathcal{C}} , \\
\end{equation}
where `$\circ$' denotes the convolution on the SK contour. The self-energy $\Sigma$ obeys
\begin{align}
    \Sigma(z_1, z_2) = & - \ii^q J^2 [G(z_1, z_2)]^{q-1} 
    \nonumber \\
    & - \ii^k (k/2)^{1/2} m \gamma^2 G(z_1, z_2)^{k-1} 
    \nonumber \\
    & \hspace{12pt} \times \biggl[\mathcal{H}(z_1, z_2) + (-1)^{k} \mathcal{H}(z_2, z_1) \biggr] ,
\end{align}
and $\mathcal{H}$ is given by
\begin{equation}
    \biggl[1 + \frac{\ii^k \gamma^2}{(2k)^{1/2}} B \biggr] \circ \mathcal{H} = \KL .
\end{equation}
These equations can be solved numerically in a similar manner as in Ref. \cite{maldacena2016}. Since the jump operators Eq.~\eqref{eq:jump_opeartors_def} are hermitian, the steady state is at infinite temperature with vanishing Keldysh Green's function $G^{\rm{K}} = G^> + G^< = 0$. Note that expanding the trace log term of the action Eq.~\eqref{eq:action_full} and keeping only the first two terms up to order $m \gamma^4$, we recover results previously reported in Refs. \cite{sa2022,kulkarni2022}. We find that the higher order terms are irrelevant, and therefore it is sufficient to keep only the first two terms. More details are found in the Supplemental Materials. It would be useful to define the effective $\mu$ coupling for generic $k$
\begin{equation} \label{eq:mu_def}
    \mu \equiv \frac{(2k)^{1/2}}{2^{k-1}} m \gamma^2 .
\end{equation}
In the small $m \gamma^2$ limit, the greater self-energy reads
$
    \Sigma^>(t) = - \ii^q J^2 [G^>(t)]^{q-1} - \ii \mu \delta(t) + O(m \gamma^4) 
$.
This is identical to the Lindblad SYK model with linear jump operators $L_{j} = \sqrt{\mu} \chi_{j}$ \cite{Garcia-Garcia:2022adg,kulkarni2022}. The solutions of the SD equations therefore agree with that case in the weak coupling limit. Moreover, the Markovian bath introduces dissipation and suppresses quantum correlations, 
giving a faster decay of the Green's function compared to the unitary dynamics
\cite{sa2022,kulkarni2022}. However, as will be demonstrated below, jump operators for $k > 2$ induce additional terms in the equations governing the OTOC, which result in an enhancement of the Lyapunov exponent with respect to the Hermitian limit.

\textit{Out-of-time-ordered correlation functions.--} Denoting $\operatorname{T}_{\mathcal{C}}$ the contour ordering in generic SK contour $\mathcal{C}$, the connected four-point function $\mathcal{F}$ is defined as
\begin{align}
    \frac{1}{N} \mathcal{F}(z_1, z_2; z_3, z_4) & = \langle \operatorname{T}_{\mathcal{C}} G(z_1, z_2) G(z_3, z_4) \rangle \nonumber\\
    & \hspace{18pt} 
    - \langle G(z_1, z_2) \rangle \langle G(z_3, z_4) \rangle .
\end{align}
The (connected) OTOCs $F_i$ are those $\mathcal{F}$ with time arguments arranged in an out-of-(real)-time ordered manner.
Denoting $z_i \equiv t_i + \tau_i$ ($\tau_i \leq 0$), for simplicity we will take $z_3 = \ii \tau_3$ and $z_4 = 0$. There are two possibilities
\begin{align}
    F_{1}(t_1, t_2) & \equiv \mathcal{F}(z_1, z_2; \ii \tau_3, 0) , \quad
    \tau_1 < \tau_3 < \tau_2 < 0  , \\
    F_{2}(t_1, t_2) & \equiv \mathcal{F}(z_1, z_2; \ii \tau_3, 0) , \quad
    \tau_2 < \tau_3 < \tau_1 < 0 . 
\end{align}
They obey the coupled recursion relations
\begin{equation} \label{eq:OTOC_recursion_relation}
    F_{i}(t_1, t_2) = \sum_{j=1}^{2} \int_{-\infty}^{\infty} \mathcal{K}_{i j}(t_1, t_2; t_3, t_4) F_{j}(t_3, t_4) \dd{t_3} \! \dd{t_4} \! ,
\end{equation}
in the long time limit $t_1, t_2 \gg t_3, t_4$. Here $\mathcal{K}_{ij}$ are the projections of the 4-pt function kernel onto the OTOC contour, whose diagonal elements are
\begin{align}
    \mathcal{K}_{11} = G^{\rm{R}}_{13} & G^{\rm{R}}_{24} 
        \biggl\{ 
            \ii^{q} (q-1) J^2 \bigl[G^>_{34}\bigr]^{q-2}
            \nonumber\\
            & + \ii^{k} (2k)^{\frac{1}{2}} (k-1) m \gamma^2 \bigl[G^>_{34}\bigr]^{k-2} \delta(t_{34}) 
            \nonumber\\
            & + (-1)^{k}(k-1) m \gamma^4 \bigl[G^>_{34}\bigr]^{k-2} \bigl[G^<_{34}\bigr]^{k}
            \nonumber\\
            & + \frac{(-1)^k}{2} k m \gamma^4 \bigl[G^>_{34}\bigr]^{2k-2}
            + O(m\gamma^6)
        \biggr\} , \\
    \mathcal{K}_{22} =  G^{\rm{R}}_{13} & G^{\rm{R}}_{24}
        \biggl\{ 
            \ii^{q} (q-1) J^2 \bigl[G^<_{34}\bigr]^{q-2}
            \nonumber\\
            & + (-1)^{k} \ii^{k} (2k)^{\frac{1}{2}} (k-1) m \gamma^2 [G^<_{34}]^{k-2} \delta(t_{34})
            \nonumber\\
            & + (-1)^k (k-1) m \gamma^4 \bigl[G^<_{34}\bigr]^{k-2} \bigl[G^>_{34}\bigr]^{k} 
            \nonumber\\
            & + \frac{(-1)^k}{2} k m \gamma^4 \bigl[G^<_{34}\bigr]^{2k-2} 
            + O(m\gamma^6)
        \biggr\} ,
\end{align}
and the off-diagonal elements are
\begin{align}
    \mathcal{K}_{12} & = - \frac{(-1)^k}{2} k m \gamma^4 G^{\rm{R}}_{14} G^{\rm{R}}_{23} \bigl[G^<_{34}\bigr]^{2k-2} + O(m\gamma^6) ,  
    \\
    \mathcal{K}_{21} & = - \frac{(-1)^k}{2} k m \gamma^4 G^{\rm{R}}_{14} G^{\rm{R}}_{23} \bigl[G^>_{34}\bigr]^{2k-2} + O(m\gamma^6) .
\end{align}
Here $G^{\rm{R}}(t, t') \equiv \theta(t-t')[G^>(t,t') - G^<(t,t')]$ denotes the retarded Green's function, and for notational simplicity we denote $G^{\rm{R}}_{ij} \equiv G^{\rm{R}}_0(t_i, t_j)$. We send all imaginary times $\tau_i$ to 0 while keeping their relative ordering fixed. Consequently, the Wightman functions with non-zero imaginary time separations reduce to the greater and lesser Green's functions \cite{liu2026}. Details of the projections are summarized in the Supplemental Materials. The Lyapunov exponent $\lambda_{\rm{L}}$ is determined by showing that the exponential growth ansatz 
\begin{equation} \label{eq:OTOC_ansatz}
    F_i(t_1, t_2) = \ee^{\lambda_{\mathrm{L}} (t_1 + t_2)/2} f_i(t_1 - t_2)
\end{equation}
solves the recursion relation Eq.~\eqref{eq:OTOC_recursion_relation}.

\begin{figure}[!t]
    \includegraphics[width=0.9\columnwidth]{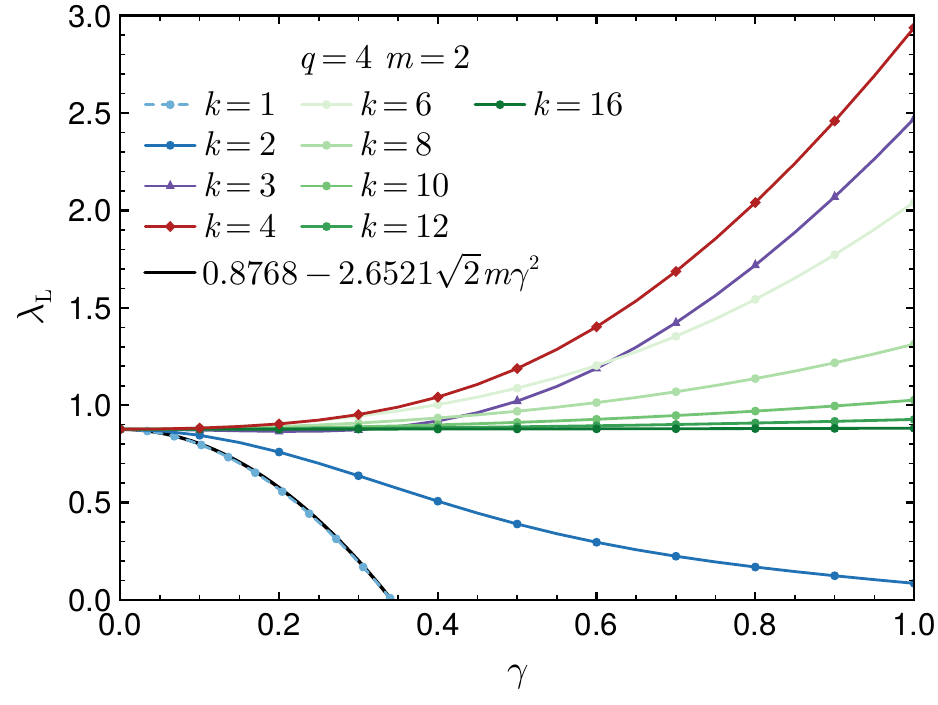}
    \caption{The Lyapunov exponent for the Lindblad SYK Eq.~\eqref{eq:Lindblad_eq} with jump operators Eq.~\eqref{eq:jump_opeartors_def} versus the dissipation strength $\gamma$ for different values of $k$ and $q=4$, $J=1$ and $m=2$. The solid black line Eq.~(\ref{eq:lambdak1}) is the best fitting of the numerical solution of $\lambda_{\mathrm{L}}$ for $k = 1$.}
    \label{fig:Lyapunov_vs_gamma}
\end{figure}
\textit{The non-integrable $q=4$ case.--} The numerical results in the case of $q=4$ for the Lyapunov exponents versus the dissipation strength $\gamma$ and different values of $k$ are presented in Fig.~\ref{fig:Lyapunov_vs_gamma}, where we set $J=1$ and $m=2$. The $k=1$ case is similar to the Lindblad SYK model with linear jump operators $L_j = \sqrt{\mu} \chi_j$ \cite{Garcia-Garcia:2024tbd}, in which case $\lambda_{\rm{L}}$ decreases monotonically with $\mu$. A numerical fitting gives $\lambda_{\rm{L}}(\mu) \approx 0.8768 - 2.6521 \mu$. Using the definition for effective $\mu$ Eq.~\eqref{eq:mu_def} we have
\begin{equation}
    \lambda_{{\rm{L}}, k=1}(\gamma) \approx 0.8768 - 2.6521 \sqrt{2} m \gamma^2 + O(m \gamma^4) \, . \label{eq:lambdak1}
\end{equation}
Results depicted in Fig.~\ref{fig:Lyapunov_vs_gamma} exhibit agreement between this analytical expression and the numerical prediction from the solution of the SD equations. On the other hand, for $k > 2$, $\lambda_{\rm{L}}$ instead increases monotonically with $\gamma$, indicating that the quantum chaotic behavior is enhanced due to the complicated interactions with the Markovian bath. As $k$ increases, however, the effective coupling $\mu$ Eq.~\eqref{eq:mu_def} decreases, and the effect of the Markovian bath becomes a perturbation. The enhancing behavior therefore is most prominent at $k=4$.

\begin{figure}[!t]
    \includegraphics[width=0.9\columnwidth]{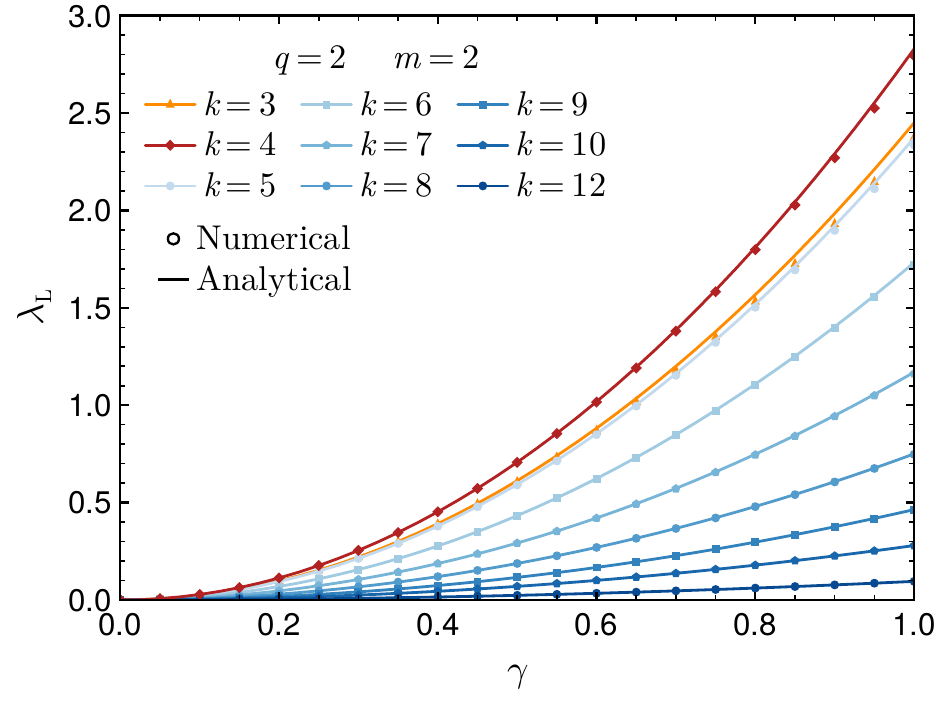}
    \caption{Lyapunov exponent $\lambda_{\mathrm{L}}$ as a function of $\gamma$ for various $k$, in the case of ($q=2$) integrable Lindblad SYK at a fixed $m=2$. $\lambda_{\mathrm{L}}$ is always positive and increases with $\gamma$ so a Markovian environment can induce quantum chaos. Scatters stand for the numerical solution of the Schwinger-Dyson equations while the lines are the analytical prediction Eqs.~\eqref{eq:mu_def} and~\eqref{eq:lamq2}.
    }
    \label{fig:lyaq2m2}
\end{figure}

\textit{Integrable $q=2$ case.--} For $q=2$, the Hermitian SYK model is integrable. Therefore, the OTOC does not experience an exponential growth and $\lambda_{\mathrm{L}} \leq 0$. After adding a Markovian bath characterized by jump operators in Eq.~(\ref{eq:jump_opeartors_def}), the Lyapunov exponent can be computed by solving the SD equations numerically along the lines of the $q = 4$ case. However, neglecting $O(m \gamma^4)$ terms, we show below that we can compute $\lambda_{\mathrm{L}}$ analytically. First, we notice that, in this limit, the SD equation 
reduces to the same equation obtained for the Lindblad SYK model with $N$ linear jump operators $L_j = \sqrt{\mu} \chi_j$ \cite{Garcia-Garcia:2022adg,Garcia-Garcia:2024tbd}
\begin{equation} \label{eq:q2_gamma2_SD}
    \Bigl[\omega + \ii \mu - J^2 G^{\rm{R}}(\omega)\Bigr] G^{\rm{R}}(\omega) = 1 .
\end{equation}
The retarded Green's function is solved by
\begin{equation} \label{eq:q2_gamma2_solution}
    G^{\rm{R}}(t) = - \ii \theta(t) \ee^{-\mu t} \frac{J_1(2 J t)}{J t} ,
\end{equation}
where $J_1(x)$ is the Bessel function of the first kind. The order $m \gamma^4$ corrections can be computed using perturbation theory. However, this leading order result already shows remarkable agreement with the numerical solutions of the SD equation, see Supplemental Materials. At leading order in $m \gamma^2$, the kernel matrix is diagonal, giving the iteration relation for the relative-time mode $f$ Eq.~\eqref{eq:OTOC_ansatz}
\begin{equation}
    f(t)
    = - J^2 \int A(t - t'; \lambda_{\rm{L}}) f(t') \dd{t'} 
    - (2k-1) \mu A(t; \lambda_{\rm{L}}) f(0) , 
\end{equation}
where
\begin{equation}
    A(t; \lambda_{\rm{L}}) = \int G^{\rm{R}}\biggl(u + \frac{1}{2} t\biggr) G^{\rm{R}}\biggl(u - \frac{1}{2} t\biggr) \ee^{-\lambda_{\rm{L}} u} \dd{u} .
\end{equation}
In frequency space,
\begin{equation}
    f(\omega) = - \frac{(2k-1) \mu A(\omega; \lambda_{\rm{L}})}{1 + J^2 A(\omega; \lambda_{\rm{L}})} \int_{-\infty}^{\infty} f(\omega') \frac{\dd{\omega'}}{2\pi} .
\end{equation}
Integrating both sides over $\omega$ gives the consistency constraint for the Lyapunov exponent $\lambda_{\rm{L}}$
\begin{equation} \label{eq:q2_gamma2_lambda_constraint}
    - (2k-1) \mu \int_{-\infty}^{\infty} \frac{A(\omega; \lambda_{\rm{L}})}{1 + J^2 A(\omega; \lambda_{\rm{L}})} \frac{\dd{\omega}}{2\pi} = 1 .
\end{equation}
At the same time, using the SD equation \eqref{eq:q2_gamma2_SD} we find the relation in frequency space
\begin{equation}
    \frac{A(\omega; \lambda_{\rm{L}})}{1 + J^2 A(\omega, \lambda_{\rm{L}})} = \frac{1}{\mu + \frac{\lambda_{\rm{L}}}{2}} \Im G^{\rm{R}}\biggl(\omega + \ii \frac{\lambda_{\rm_{L}}}{2}\biggr) .
\end{equation}
Substituting it into Eq.~\eqref{eq:q2_gamma2_lambda_constraint} and using the normalization condition $- \pi^{-1} \int \Im(G^{\rm{R}}) \dd{\omega} = 1$, we obtain
\begin{equation}
    \lambda_{\rm{L}} = 2(k-2) \mu + O(m \gamma^4),
    \label{eq:lamq2}
\end{equation}
with $\mu$ given by Eq.~(\ref{eq:mu_def}).
Numerical results, depicted in Fig.~\ref{fig:lyaq2m2}, show an excellent agreement with this simple analytical prediction. The Lyapunov exponent is positive and increases with $\gamma^2$. 
This provides conclusive evidence that a Markovian bath can induce quantum chaos provided that the jump operators are of the form Eq.~(\ref{eq:jump_opeartors_def}) with $k > 2$. Moreover, in this perturbative region where only $m\gamma^2$ terms are kept, the Lyapunov exponent increases linearly with $\mu$. 

\begin{figure}[!t]
    \includegraphics[width=0.9\columnwidth]{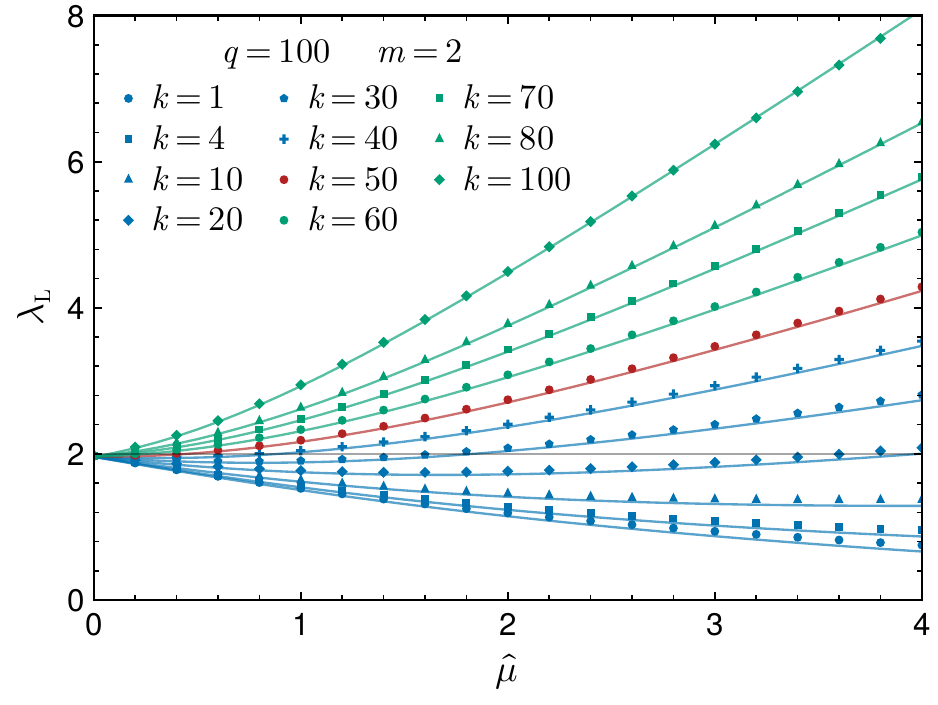}
    \caption{Lyapunov exponents for $q=100$ for various $k$ and $\mathcal{J}=1$. Scatters stand for the numerical solution of the Schwinger-Dyson equation while the lines are the analytical prediction Eq.~(\ref{eq:lambda_L_large_q}) that includes the leading $1/q$ correction.}
    \label{fig:large_q}
\end{figure}

\textit{Large $q$ limit.--} 
In the large $q$ and large $k$ limit, with fixed $k / q$, we have the $O(1)$ couplings
\begin{equation}
    \mathcal{J}^2 \equiv \frac{q J^2}{2^{q-1}} , 
    \quad
    \hat{\mu} 
    \equiv \frac{q (2k)^{1/2} m \gamma^2}{2^{k-1}} ,
    \quad
    \rho \equiv \frac{k - 1}{q} .
\end{equation}
Due to this scaling limit, we may consider only the contact term and neglect higher order contributions in $\gamma^2$, as in the $q=2$ case. We can make an ansatz $G^{\mathrm{R}}(t) = - \ii \theta(t) \ee^{g(t) / q}$. The SD equation then reduces to the same equation obtained in Refs.~\cite{Garcia-Garcia:2022adg,Garcia-Garcia:2024tbd}, and the solution is given by
$
    \ee^{g(t)} = \alpha^2 / \mathcal{J}^2 \cosh^2(\alpha |t| + \gamma)
$,
with
$
    \alpha^2 = \mathcal{J}^2 +\hat{\mu}^2 / 4 
$
and
$
    \tanh(\gamma) = \hat{\mu} / 2 \alpha
$.
The Green's function therefore has a decay rate $2 \alpha / q$.
Acting both sides of the OTOC recursion equation \eqref{eq:OTOC_recursion_relation} with $\partial_{t_1} \partial_{t_2}$, and inserting the exponential growth ansatz Eq.~\eqref{eq:OTOC_ansatz}, 
\begin{equation} \label{eq:large_q_gamma2_f_equation}
    - f''(t)- \bigg[
    \frac{2 \alpha^2}{\cosh^2(\alpha |t| + \gamma)} + 2 \rho \hat{\mu} \delta(t)
    \bigg] f(t) = - \frac{\lambda_{\rm{L}}^2}{4} f(t) .
\end{equation}
This equation is symmetric under $t \leftrightarrow - t$, and for $t \neq 0$ has the following solution
\begin{equation} \label{eq:large_q_f_t}
    f(t) = \ee^{- \frac{\lambda_{\rm{L}}}{2} |t|} \bigg[
        \frac{\lambda_{\rm{L}}}{2} + \alpha \tanh(\alpha |t| + \gamma)
    \bigg] , 
    \quad 
    (t \neq 0) .
\end{equation}
Integrating both sides of Eq.~\eqref{eq:large_q_gamma2_f_equation} about $t = 0$ gives the constraint
$
    f'(0^+) - f'(0^-) + 2 \rho \hat{\mu} f(0) = 0 .
$
Substitution into the solution Eq.~\eqref{eq:large_q_f_t} gives a quadratic equation for the Lyapunov exponent, whose solution is
\begin{equation} \label{eq:lambda_L_large_q}
    \lambda_{\rm{L}} = \biggl(\rho - \frac{1}{2} \biggr)\hat{\mu} + \Biggl[
        \biggl(\rho + \frac{1}{2}\biggr)^2 \hat{\mu}^2 + 4 \mathcal{J}^2
    \Biggr]^{\frac{1}{2}} - \frac{4 \alpha}{q} .
\end{equation}
Results depicted in Fig.~\ref{fig:large_q} show excellent agreement between these large $q$  analytical results and those coming from the numerical solution of the SD equations for $q \gg 1$. As a sanity check, in the case $k \ll q$ we have $\rho = 0$, giving
$
    \lambda_{\mathrm{L}} = - \frac{1}{2} \hat{\mu} + \sqrt{\hat{\mu}^2/4 + 4 \mathcal{J}^2} \leq 2 \mathcal{J} ,
$
such that the dissipation term suppresses quantum chaos, which reproduces the results obtained in~\cite{Garcia-Garcia:2024tbd}. 
For $\rho \geq 1/2$, the Lyapunov exponent instead increases monotonically with $\hat{\mu}$. 
On the other hand, for $0 < \rho < 1/2$ (equivalently $ 1 \ll k < q/2$), the Lyapunov exponent exhibits a non-monotonic behavior with respect to $\hat{\mu}$, which experiences three phases: weakening, recovering, and enhancing. For convenience let us denote 
$
    \hat{\mu}_1 \equiv \frac{\sqrt{2} (1-2\rho)}{\sqrt{\rho}(1 + 2\rho)} \mathcal{J}
$
and
$
    \hat{\mu}_2 \equiv \frac{1 - 2 \rho}{\rho} \mathcal{J}
$.
At the first weakening phase 
$
    0 \leq \hat{\mu} \leq \hat{\mu}_1
$,
$\lambda_{\mathrm{L}}$ decreases from $2\mathcal{J}$ down to the minimum
$
    4 \mathcal{J} \sqrt{2 \rho} / (2\rho + 1) .
$
Then it increases monotonically in the second phase, $\hat{\mu}_1 < \hat{\mu} \leq \hat{\mu}_2$, until it recovers the unitary value $2 \mathcal{J}$.
After that $\lambda_{\mathrm{L}}$ keeps increasing in the enhancing phase $\hat{\mu} > \hat{\mu}_2$.

\textit{Conclusions and discussions.--} In this work we investigate the dynamics of the SYK$_q$ model coupled to a Markovian bath characterized by isotropic random $k$-body Majoranas. 
The effect of the Markovian reservoir is twofold. On the one hand, it introduces dissipation and suppresses quantum correlations, such that the decay rate of the system Green's function increases monotonically with the system-bath coupling in the case of $q=2$ and in the large $q$ limit as well. On the other hand, for $k \geq 2$ it also generates non-trivial contributions to the OTOC kernel and therefore enhances or induces, if the Hermitian system was integrable, quantum chaos. We demonstrate this phenomenon analytically and numerically, by solving explicitly the SD equations, in both integrable and non-integrable cases. 
The $q = 2$ SYK is quadratic in the fermionic fields so the Lyapunov exponent cannot be positive because it is an integrable system. When coupled to the Markovian bath, this is still true \cite{garcia2025} if the jump operators are linear in the Majorana fields. However, for the more general operators of Eq.~\eqref{eq:jump_opeartors_def} with $k>2$, the system develops a positive Lyapunov exponent which is a signature of quantum chaotic behavior. Our results challenge the common belief that quantum chaos is always suppressed by the environment by providing a concrete analytically tractable example where an environment induces many-body quantum chaos.

An interesting question concerns whether the enhancement of scrambling by turning on a Markovian bath, described in the paper by a larger Lyapunov exponent, occurs in other systems and impacts other observables such as the operator size growth \cite{Roberts:2018mnp,Liu:2024stj}. We wish to address this question in the near future.

\textit{Note Added.--} While finishing the writing of this manuscript, we became aware of the preprint~\cite{pelliconi2026} that studies the dynamics of the SYK model coupled to a dissipative cavity. Although the model is different, it is also identified a range of parameters for which the Lyapunov exponent is enhanced as the dissipation strength increases. A later preprint \cite{pacchioni2026} reaches similar conclusions, still in the context of the SYK in a dissipative cavity, but their observable is not the OTOC but the spectral form factor.

We were partially supported by the National Science Foundation of China
(NSFC), Individual Grant No.12374138.

\bibliography{references}

@misc{pacchioni2026,
	title={Dissipation-induced Sachdev-Ye-Kitaev physics in many-body cavity quantum electrodynamics}, 
	author={Pietro Pacchioni and Filippo Ferrari and Vincenzo Savona and Matteo Seclì},
	year={2026},
	eprint={2608.23557},
	archivePrefix={arXiv},
	primaryClass={quant-ph},
	url={https://arxiv.org/abs/2608.23557}, 
}

@misc{pengfei2026,
	title={Enhancing Many-Body Chaos via Entropy Injection from Environment}, 
	author={Yuke Zhang and Wenbo Zhou and Pengfei Zhang},
	year={2026},
	eprint={2606.11784},
	archivePrefix={arXiv},
	primaryClass={quant-ph},
	url={https://arxiv.org/abs/2606.11784}, 
}

@misc{pelliconi2026,
	title={Dissipation-enhanced scrambling in the SYK model coupled to a lossy cavity}, 
	author={Pietro Pelliconi and Bastien Lapierre and Shinsei Ryu},
	year={2026},
	eprint={2608.19310},
	archivePrefix={arXiv},
	primaryClass={quant-ph},
	url={https://arxiv.org/abs/2608.19310}, 
}

@article{altman2023,
  author    = {Weinstein, Zack and Kelly, Shane P. and Marino, Jamir and Altman, Ehud},
  doi       = {10.1103/PhysRevLett.131.220404},
  issue     = {22},
  journal   = {Phys. Rev. Lett.},
  month     = {Nov},
  numpages  = {7},
  pages     = {220404},
  publisher = {American Physical Society},
  title     = {{Scrambling Transition in a Radiative Random Unitary Circuit}},
  url       = {https://link.aps.org/doi/10.1103/PhysRevLett.131.220404},
  volume    = {131},
  year      = {2023}
}

@article{benet2001,
  author  = {L. Benet and T. Rupp and H. A. Weidenm{\"u}ller},
  journal = {Phys. Rev. Lett.},
  month   = jun,
  pages   = {010601},
  title   = {Nonuniversal behavior of the $k$-body embedded Gaussian unitary ensemble of random matrices},
  volume  = {87},
  year    = {2001}
}

@article{bergamasco2023,
  author    = {Bergamasco, Pablo D. and Carlo, Gabriel G. and Rivas, Alejandro M. F.},
  doi       = {10.1103/PhysRevE.108.024208},
  issue     = {2},
  journal   = {Phys. Rev. E},
  month     = {Aug},
  numpages  = {7},
  pages     = {024208},
  publisher = {American Physical Society},
  title     = {{Quantum Lyapunov exponent in dissipative systems}},
  url       = {https://link.aps.org/doi/10.1103/PhysRevE.108.024208},
  volume    = {108},
  year      = {2023}
}

@article{berman1978,
  author  = {G.P. Berman and G.M. Zaslavsky},
  doi     = {http://dx.doi.org/10.1016/0378-4371(78)90190-5},
  issn    = {0378-4371},
  journal = {Physica A},
  number  = {3},
  pages   = {450 - 460},
  title   = {{Condition of stochasticity in quantum nonlinear systems}},
  url     = {http://www.sciencedirect.com/science/article/pii/0378437178901905},
  volume  = {91},
  year    = {1978}
}

@article{bohigas1971,
  author  = {O. Bohigas and J. Flores},
  journal = {Phys. Lett. B},
  number  = {4},
  pages   = {261--263},
  title   = {Two-body random Hamiltonian and level density},
  volume  = {34},
  year    = {1971}
}

@article{bohigas1984,
  author    = {Bohigas, O. and Giannoni, M. J. and Schmit, C.},
  doi       = {10.1103/PhysRevLett.52.1},
  issue     = {1},
  journal   = {Phys. Rev. Lett.},
  month     = {Jan},
  numpages  = {0},
  pages     = {1--4},
  publisher = {American Physical Society},
  title     = {{Characterization of Chaotic Quantum Spectra and Universality of Level Fluctuation Laws}},
  url       = {https://link.aps.org/doi/10.1103/PhysRevLett.52.1},
  volume    = {52},
  year      = {1984}
}

@article{chen2017a,
  author    = {Chen, Yiming and Zhai, Hui and Zhang, Pengfei},
  doi       = {10.1007/jhep07(2017)150},
  issn      = {1029-8479},
  journal   = {Journal of High Energy Physics},
  month     = {Jul},
  number    = {7},
  publisher = {Springer Science and Business Media LLC},
  title     = {Tunable quantum chaos in the Sachdev-Ye-Kitaev model coupled to a thermal bath},
  url       = {http://dx.doi.org/10.1007/JHEP07(2017)150},
  volume    = {2017},
  year      = {2017}
}

@article{Ferrari:2019ogc,
  archiveprefix = {arXiv},
  author        = {Ferrari, Frank and Schaposnik Massolo, Fidel I.},
  doi           = {10.1103/PhysRevD.100.026007},
  eprint        = {1903.06633},
  journal       = {Phys. Rev. D},
  number        = {2},
  pages         = {026007},
  primaryclass  = {hep-th},
  title         = {{Phases Of Melonic Quantum Mechanics}},
  volume        = {100},
  year          = {2019}
}

@article{french1970,
  author  = {J. B. French and S. S. M. Wong},
  journal = {Phys. Lett. B},
  number  = {7},
  pages   = {449--452},
  title   = {Validity of random matrix theories for many-particle systems},
  volume  = {33},
  year    = {1970}
}

@article{Garcia-Garcia:2022adg,
  archiveprefix = {arXiv},
  author        = {Garc{\'\i}a-Garc{\'\i}a, Antonio M. and S{\'a}, Lucas and Verbaarschot, Jacobus J. M. and Zheng, Jie Ping},
  doi           = {10.1103/PhysRevD.107.106006},
  eprint        = {2210.01695},
  journal       = {Phys. Rev. D},
  number        = {10},
  pages         = {106006},
  primaryclass  = {hep-th},
  title         = {{Keldysh wormholes and anomalous relaxation in the dissipative Sachdev-Ye-Kitaev model}},
  volume        = {107},
  year          = {2023}
}

@article{Garcia-Garcia:2024tbd,
  archiveprefix = {arXiv},
  author        = {Garc{\'\i}a-Garc{\'\i}a, Antonio M. and Verbaarschot, Jacobus J. M. and Zheng, Jie-ping},
  doi           = {10.1103/PhysRevD.110.086010},
  eprint        = {2403.12359},
  journal       = {Phys. Rev. D},
  number        = {8},
  pages         = {086010},
  primaryclass  = {hep-th},
  title         = {{Lyapunov exponent as a signature of dissipative many-body quantum chaos}},
  volume        = {110},
  year          = {2024}
}

@article{garcia2016,
  archiveprefix = {arXiv},
  author        = {Garc\'{\i}a-Garc\'{\i}a, Antonio M. and Verbaarschot, Jacobus J. M.},
  doi           = {10.1103/PhysRevD.94.126010},
  eprint        = {1610.03816},
  issue         = {12},
  journal       = {Phys. Rev. D},
  month         = {Dec},
  numpages      = {13},
  pages         = {126010},
  primaryclass  = {hep-th},
  publisher     = {American Physical Society},
  title         = {{Spectral and thermodynamic properties of the Sachdev-Ye-Kitaev model}},
  url           = {https://link.aps.org/doi/10.1103/PhysRevD.94.126010},
  volume        = {94},
  year          = {2016}
}

@article{garcia2023,
  author  = {Antonio M. Garc{\'i}a-Garc{\'i}a and Lucas S{\'a} and Jacobus J. M. Verbaarschot and Jie Ping Zheng},
  journal = {Phys. Rev. D},
  month   = may,
  pages   = {106006},
  title   = {Keldysh wormholes and anomalous relaxation in the dissipative Sachdev-Ye-Kitaev model},
  volume  = {107},
  year    = {2023}
}

@article{garcia2024,
  author  = {Antonio M. Garc{\'i}a-Garc{\'i}a and Jacobus J. M. Verbaarschot and Jie-Ping Zheng},
  journal = {Phys. Rev. D},
  month   = oct,
  pages   = {086010},
  title   = {Lyapunov exponent as a signature of dissipative many-body quantum chaos},
  volume  = {110},
  year    = {2024}
}

@article{garcia2024d,
  author    = {Garc\'{\i}a-Garc\'{\i}a, Antonio M. and Liu, Chang and Verbaarschot, Jacobus J. M.},
  doi       = {10.1103/PhysRevLett.133.091602},
  issue     = {9},
  journal   = {Phys. Rev. Lett.},
  month     = {Aug},
  numpages  = {6},
  pages     = {091602},
  publisher = {American Physical Society},
  title     = {Sparsity-Independent Lyapunov Exponent in the Sachdev-Ye-Kitaev Model},
  url       = {https://link.aps.org/doi/10.1103/PhysRevLett.133.091602},
  volume    = {133},
  year      = {2024}
}

@misc{garcia2025,
  author = {Antonio M. Garc{\'i}a-Garc{\'i}a and Chang Liu and Lucas S{\'a} and Jacobus J. M. Verbaarschot and Jie Ping Zheng},
  note   = {preprint},
  title  = {Anatomy of information scrambling and decoherence in the integrable Sachdev-Ye-Kitaev model},
  year   = {2025}
}

@misc{garcia2026,
  archiveprefix = {arXiv},
  author        = {Antonio M. García-García and Lucas Sá and Jacobus J. M. Verbaarschot and Jie-Ping Zheng},
  eprint        = {2605.27512},
  primaryclass  = {hep-th},
  title         = {Many-Body Quantum Chaos At All Time Scales},
  url           = {https://arxiv.org/abs/2605.27512},
  year          = {2026}
}

@article{gu2022,
  author    = {Gu, Yingfei and Kitaev, Alexei and Zhang, Pengfei},
  doi       = {10.1007/jhep03(2022)133},
  issn      = {1029-8479},
  journal   = {Journal of High Energy Physics},
  month     = Mar,
  number    = {3},
  publisher = {Springer Science and Business Media LLC},
  title     = {A two-way approach to out-of-time-order correlators},
  url       = {http://dx.doi.org/10.1007/JHEP03(2022)133},
  volume    = {2022},
  year      = {2022}
}

@book{kamenev2023field,
  author    = {Kamenev, A.},
  isbn      = {9781108488259},
  publisher = {Cambridge University Press},
  title     = {Field Theory of Non-Equilibrium Systems},
  url       = {https://books.google.com/books?id=y9WgEAAAQBAJ},
  year      = {2023}
}

@misc{kitaev2015,
  author       = {Alexander Kitaev},
  howpublished = {Talks at KITP: String seminar and Entanglement 2015 program, 12 Feb, 7 Apr, 27 May},
  note         = {\url{http://online.kitp.ucsb.edu/online/entangled15/}},
  title        = {A simple model of quantum holography},
  year         = {2015}
}

@article{kobrin2020,
  author    = {Kobrin, Bryce and Yang, Zhenbin and Kahanamoku-Meyer, Gregory D. and Olund, Christopher T. and Moore, Joel E. and Stanford, Douglas and Yao, Norman Y.},
  doi       = {10.1103/PhysRevLett.126.030602},
  issue     = {3},
  journal   = {Phys. Rev. Lett.},
  month     = {Jan},
  numpages  = {6},
  pages     = {030602},
  publisher = {https://link.aps.org/doi/10.1103/PhysRevB.102.224305},
  title     = {Many-Body Chaos in the {Sachdev-Ye-Kitaev Model}},
  url       = {https://link.aps.org/doi/10.1103/PhysRevLett.126.030602},
  volume    = {126},
  year      = {2021}
}

@article{kulkarni2022,
  author  = {Anish Kulkarni and Tokiro Numasawa and Shinsei Ryu},
  journal = {Phys. Rev. B},
  month   = aug,
  pages   = {075138},
  title   = {Lindbladian dynamics of the Sachdev-Ye-Kitaev model},
  volume  = {106},
  year    = {2022}
}

@article{larkin1969,
  author  = {Larkin, A I and Ovchinnikov, Yu N},
  journal = {Sov. Phys. JETP},
  number  = {6},
  pages   = {1200--1205},
  title   = {{Quasiclassical method in the theory of superconductivity}},
  url     = {http://jetp.ras.ru/cgi-bin/e/index/e/28/6/p1200?a=list},
  volume  = {28},
  year    = {1969}
}

@article{lindblad1976,
  author  = {Goran Lindblad},
  journal = {Commun. Math. Phys.},
  number  = {2},
  pages   = {119--130},
  title   = {On the generators of quantum dynamical semigroups},
  volume  = {48},
  year    = {1976}
}

@article{Liu:2024stj,
  archiveprefix = {arXiv},
  author        = {Liu, Jiasheng and Meyer, Rene and Xian, Zhuo-Yu},
  doi           = {10.1007/JHEP08(2024)092},
  eprint        = {2403.07115},
  journal       = {JHEP},
  pages         = {092},
  primaryclass  = {hep-th},
  title         = {{Operator size growth in Lindbladian SYK}},
  volume        = {08},
  year          = {2024}
}

@misc{liu2026,
  archiveprefix = {arXiv},
  author        = {Xianlong Liu and Jie-ping Zheng and Antonio M. García-García},
  eprint        = {2602.02750},
  primaryclass  = {quant-ph},
  title         = {Inducing, and enhancing, many-body quantum chaos by continuous monitoring},
  url           = {https://arxiv.org/abs/2602.02750},
  year          = {2026}
}

@article{maldacena2015,
  archiveprefix = {arXiv},
  author        = {Maldacena, Juan and Shenker, Stephen H. and Stanford, Douglas},
  eprint        = {1503.01409},
  journal       = {J. High Energy Phys.},
  number        = {106},
  primaryclass  = {hep-th},
  title         = {A bound on chaos},
  url           = {http://dx.doi.org/10.1007/JHEP08(2016)106},
  volume        = {2016},
  year          = {2016}
}

@article{maldacena2016,
  author  = {Juan Maldacena and Douglas Stanford},
  journal = {Phys. Rev. D},
  month   = nov,
  pages   = {106002},
  title   = {Remarks on the Sachdev-Ye-Kitaev model},
  volume  = {94},
  year    = {2016}
}

@article{pengfei2023,
  author    = {Zhang, Pengfei and Yu, Zhenhua},
  doi       = {10.1103/PhysRevLett.130.250401},
  issue     = {25},
  journal   = {Phys. Rev. Lett.},
  month     = {Jun},
  numpages  = {7},
  pages     = {250401},
  publisher = {American Physical Society},
  title     = {Dynamical Transition of Operator Size Growth in Quantum Systems Embedded in an Environment},
  url       = {https://link.aps.org/doi/10.1103/PhysRevLett.130.250401},
  volume    = {130},
  year      = {2023}
}

@article{Roberts:2018mnp,
  archiveprefix = {arXiv},
  author        = {Roberts, Daniel A. and Stanford, Douglas and Streicher, Alexandre},
  doi           = {10.1007/JHEP06(2018)122},
  eprint        = {1802.02633},
  journal       = {JHEP},
  pages         = {122},
  primaryclass  = {hep-th},
  title         = {{Operator growth in the SYK model}},
  volume        = {06},
  year          = {2018}
}

@article{sa2022,
  author  = {Lucas S{\'a} and Pedro Ribeiro and Toma{\v{z}} Prosen},
  journal = {Phys. Rev. Research},
  month   = jun,
  pages   = {L022068},
  title   = {Lindbladian dissipation of strongly-correlated quantum matter},
  volume  = {4},
  year    = {2022}
}

@article{sachdev1993,
  author  = {Subir Sachdev and Jinwu Ye},
  journal = {Phys. Rev. Lett.},
  month   = may,
  pages   = {3339--3342},
  title   = {Gapless spin-fluid ground state in a random quantum Heisenberg magnet},
  volume  = {70},
  year    = {1993}
}

@article{syzranov2018,
  author    = {Syzranov, S. V. and Gorshkov, A. V. and Galitski, V.},
  doi       = {10.1103/PhysRevB.97.161114},
  issue     = {16},
  journal   = {Phys. Rev. B},
  month     = {Apr},
  numpages  = {5},
  pages     = {161114},
  publisher = {American Physical Society},
  title     = {Out-of-time-order correlators in finite open systems},
  url       = {https://link.aps.org/doi/10.1103/PhysRevB.97.161114},
  volume    = {97},
  year      = {2018}
}

@article{tuziemski2019,
  author    = {Tuziemski, Jan},
  doi       = {10.1103/PhysRevA.100.062106},
  issue     = {6},
  journal   = {Phys. Rev. A},
  month     = {Dec},
  numpages  = {10},
  pages     = {062106},
  publisher = {American Physical Society},
  title     = {{Out-of-time-ordered correlation functions in open systems: A Feynman-Vernon influence functional approach}},
  url       = {https://link.aps.org/doi/10.1103/PhysRevA.100.062106},
  volume    = {100},
  year      = {2019}
}

@article{yoshida2019,
  author    = {Yoshida, Beni and Yao, Norman Y.},
  doi       = {10.1103/PhysRevX.9.011006},
  issue     = {1},
  journal   = {Phys. Rev. X},
  month     = {Jan},
  numpages  = {17},
  pages     = {011006},
  publisher = {American Physical Society},
  title     = {{Disentangling Scrambling and Decoherence via Quantum Teleportation}},
  url       = {https://link.aps.org/doi/10.1103/PhysRevX.9.011006},
  volume    = {9},
  year      = {2019}
}

@article{zanardi2021,
  author    = {Zanardi, Paolo and Anand, Namit},
  doi       = {10.1103/PhysRevA.103.062214},
  issue     = {6},
  journal   = {Phys. Rev. A},
  month     = {Jun},
  numpages  = {16},
  pages     = {062214},
  publisher = {American Physical Society},
  title     = {{Information scrambling and chaos in open quantum systems}},
  url       = {https://link.aps.org/doi/10.1103/PhysRevA.103.062214},
  volume    = {103},
  year      = {2021}
}

\onecolumngrid
\clearpage

\setcounter{table}{0}
\renewcommand{\thetable}{S\arabic{table}}%
\setcounter{figure}{0}
\renewcommand{\thefigure}{SM\arabic{figure}}%
\setcounter{equation}{0}
\renewcommand{\theequation}{S\arabic{equation}}%
\setcounter{page}{1}
\renewcommand{\thepage}{SM-\arabic{page}}%
\setcounter{secnumdepth}{3}
\setcounter{section}{0}
\renewcommand{\thesection}{S\arabic{section}}%
\setcounter{subsection}{0}
\renewcommand{\thesubsection}{\thesection.\arabic{subsection}}

\begin{center}
        \textbf{Supplemental Materials for }
		\textbf{``From integrability to many-body quantum chaos through \\ a Markovian bath''}
        \\
		\vspace{10pt}
        Xianlong Liu and Antonio M. Garc\'ia-Garc\'ia
\end{center}

\tableofcontents

\section{Lindbladian Path integral}
\label{appendix:lindbladian_path_integral}

\subsection{Majorana operator basis}
\label{appendix_sub:Majorana_operator_basis}

For later convenience we first summarize the operator basis consisting of $N$ Majorana fermions $\chi_j \, (j=1,\dots, N)$ with anti-commutation relations $\{\chi_i, \chi_j\} = \delta_{ij}$. Each Majorana fermion can be represented through the Jordan-Wigner transformation as a traceless Hermitian matrix of size $2^{\lceil N/2 \rceil}$ satisfying $\chi_j^2 = 1/2$. Denoting $I$ a set of indices $I \equiv \{j_1, j_2, \dots , j_{|I|} \mid 1 \leq j_1 < j_2 < \dots j_{|I|} \leq N \}$, one can construct an operator basis
\begin{equation}
 \Bigl\{
        \Gamma_{I} \Bigm| \Gamma_{I} \equiv \ii^{\frac{|I|(|I|-1)}{2}} \chi_{i_1} \chi_{i_2} \dots \chi_{i_{|I|}} 
    \Bigr\} \, .
\end{equation}
By construction there are several properties of these operators
\begin{itemize}
    \item There are $2^N$ such operators.
    \item $\Gamma_I$'s are also Hermitian and traceless:
        \begin{equation}
        \Gamma_{I}^{\dagger} = \Gamma_{I} \, , 
        \quad
        \tr(\Gamma_{I}) = 0 \, .
        \end{equation}
    \item Since $\Gamma_{I}^{\dagger} \Gamma_{I}^{{}} = 2^{-|I|} \mathbb{I}$,
        \begin{equation}
        \tr\bigl(\Gamma_{I}^{\dagger} \Gamma_{J}^{{}}\bigr) = \delta_{I J} 2^{-|I|} \tr(\mathbb{I}) \, .
        \end{equation}
    \item They obey the commutation relation
        \begin{equation}
        \Gamma_{I} \Gamma_{J} = (-1)^{|I| |J| - \left|I \cap J \right|} \Gamma_{J} \Gamma_{I} \, ,
        \end{equation}
        where $\left|I \cap J \right|$ denotes the number of indices shared by the index set $I$ and $J$. 
\end{itemize}
Since $\{\Gamma_I\}$ forms an orthonormal basis, any operator $\mathcal{O}$ constructed by the Majorana fermions can be expanded as
\begin{equation}
\mathcal{O}[\chi] = \sum_{I} c_{I} \Gamma_{I} \, , 
\end{equation}
where the coefficients can be computed by
\begin{equation}
c_{I} = 2^{|I| - \lceil N/2 \rceil} \tr\bigl(\Gamma_{I}^{\dagger} \mathcal{O} \bigr) \, .
\end{equation}

Introducing the bi-local collective field on the Schwinger-Keldysh contour
\begin{equation}
G(z, z') = - \frac{\ii}{N} \sum_{i=1}^{N} \chi_i(z) \chi_i(z') \, ,
\end{equation}
in the large $N$ limit we have the relation 
\begin{equation}
    \sum_{|I|=q} \Gamma_{I}(z) \Gamma_{I}(z') = \frac{\ii^q N^q}{q!} G(z, z')^q + \dots
\end{equation}

\subsection{General formulation of Schwinger-Keldysh path integral for Lindblad dynamics}

\begin{figure}[!thb]
\begin{center}
\includegraphics[width=0.5\textwidth]{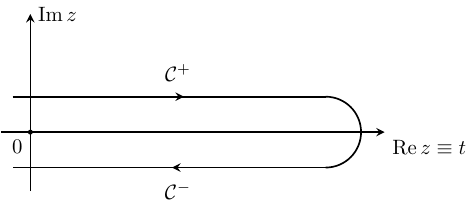}
\caption{Schwinger-Keldysh contour.}
\end{center}
\end{figure}

We first review the Schwinger-Keldysh path integral for generic Lindblad dynamics of $N$ Majorana fermions~\cite{kamenev2023field,sa2022}. In order to derive a coherent state path integral, we first introduce complex fermions $c_j$, $c_j^\dagger$ and write
\begin{equation}
    \chi_{2j-1}=\frac{1}{\sqrt{2}}(c_j+c_j^\dagger) \, , \qquad
    \chi_{2j}=\frac{\ii}{\sqrt{2}}(c_j-c_j^\dagger) \, .
\end{equation}
The fermionic coherent state trace formula then gives
\begin{equation}
    \operatorname{tr}\bigl(\hat{\mathcal{O}}(\chi)\bigr) = \int \dd{\bar{\psi}} \dd{\psi} \ee^{-\bar{\psi} \psi} \bra{\psi} \hat{\mathcal{O}} \ket{-\psi} \, ,
\end{equation}
where the minus sign inside the ket is inserted for boundary condition of fermions. The path integral is expressed in terms of the real Grassmann fields $a_i$ associated with the Majorana fermions, which are related to the complex Grassmann fields $\psi$ and $\bar{\psi}$ by a linear transformation 
\begin{equation}
    a_{2j-1}=\frac{1}{\sqrt{2}}(\psi_j+\bar\psi_j) \, ,\qquad
    a_{2j}=\frac{\ii}{\sqrt{2}}(\psi_j-\bar\psi_j) \, .
\end{equation}

Recall that the Lindblad equation is
\begin{equation*}
    \frac{\dd \rho}{\dd t} = \mathcal{L}(\rho) = - \ii [\hat{H}, \rho] + \sum_{a} \left(2 \hat{L}_a \rho \hat{L}_a^\dagger - \{\hat{L}_a^\dagger \hat{L}_a, \rho\} \right) \, .
\end{equation*}
Employing Trotter decomposition, we have
\begin{equation}
    \rho(t) = \ee^{(t - t_0) \mathcal{L}}(\rho(t_0)) = \lim_{\mathcal{N} \rightarrow \infty} (1 + \Delta t \, \mathcal{L})^{\mathcal{N}} (\rho(t_0)) \, , 
    \quad
    \Delta t \equiv \frac{t - t_0}{\mathcal{N}} \, .
\end{equation}
For each infinitesimal step we have the matrix element
\begin{equation}
    \bra{\psi_{n+1}^+} \rho(t_{n+1}) \ket{-\psi_{n+1}^-} = \bra{\psi_{n+1}} (1 + \Delta t \, \mathcal{L})(\rho(t_n)) \ket{-\psi_{n+1}} \, .
\end{equation}
We can insert the identity operators on both sides of $\rho(t_n)$ to get
\begin{align}
& \bra{\psi_{n+1}^+} \rho(t_{n+1}) \ket{-\psi_{n+1}^-} \nonumber \\
= & \int \dd{\bar{\psi}_{n}^+} \dd{\psi_{n}^{+}} \dd{\bar{\psi}_{n}^-} \dd{\psi_{n}^{-}} \, \ee^{-\bar{\psi}_{n}^+ \psi_{n}^+} \ee^{-\bar{\psi}_{n}^- \psi_{n}^{-}} \, \bra{\psi_{n+1}^+}  
(1 + \Delta t \, \mathcal{L}) 
\biggl( 
    \dyad{\psi_{n}^+}{\psi_n^+}  \rho(t_n) \dyad{-\psi_n^-}{-\psi_n^-} 
\biggr) \ket{-\psi_{n+1}^{-}} \nonumber \\
= & \int \dd{\bar{\psi}_{n}^+} \dd{\psi_{n}^{+}} \dd{\bar{\psi}_{n}^-} \dd{\psi_{n}^{-}} \, \ee^{-\bar{\psi}_{n}^+ \psi_{n}^+} \ee^{-\bar{\psi}_{n}^- \psi_{n}^{-}} \, \bra{\psi_{n+1}^+}  
(1 + \Delta t \, \mathcal{L}) \Bigl( \dyad{\psi_{n}^+}{-\psi_n^-} \Bigr) \ket{-\psi_{n+1}^-} \bra{\psi_n^+}  \rho(t_n) \ket{-\psi_n^-}\, . \label{eq:rho_matrix_element}
\end{align}
Using the definition of $\mathcal{L}$, we find
\begin{align}
\mathcal{L}(\dyad{\psi_{n}^+}{-\psi_{n}^{-}}) = &\,
- \ii \hat{H} \dyad{\psi_{n}^+}{-\psi_{n}^{-}}
+ \dyad{\psi_{n}^+}{-\psi_{n}^{-}} (\ii \hat{H}) \nonumber \\
&
+ \sum_{a} 
\biggl(
2\hat{L}_a  \dyad{\psi_{n}^+}{-\psi_{n}^{-}} \hat{L}_a^\dagger
- \hat{L}_a^\dagger \hat{L}_{a} \dyad{\psi_{n}^+}{-\psi_{n}^{-}} 
-  \dyad{\psi_{n}^+}{-\psi_{n}^{-}} \hat{L}_a^\dagger  \hat{L}_{a} 
\biggr) \, .
\end{align}
Denoting
\begin{align}
    L_a^+[\bar{\psi}, \psi] & \equiv \frac{\bra{\psi_{n+1}^+} \hat{L}_a \ket{\psi_n^+}}{\braket{\psi_{n+1}^+}{\psi_n^+}} \, ,
    \quad
    & \bar{L}_a^-[-\bar{\psi}, -\psi] & \equiv \frac{\bra{-\psi_{n}^-} \hat{L}_a^{\dagger} \ket{-\psi_{n+1}^-}}{\braket{-\psi_{n}^-}{-\psi_{n+1}^-}} \, , \\
    (\bar{L}_a L_a)^+[\bar{\psi}, \psi] & \equiv \frac{\bra{\psi_{n+1}^+} \hat{L}_a^\dagger \hat{L}_a \ket{\psi_n^+}}{\braket{\psi_{n+1}^+}{\psi_n^+}} \, ,
    \quad
    & (\bar{L}_a L_a)^-[-\bar{\psi}, -\psi] & \equiv \frac{\bra{-\psi_{n}^-} \hat{L}_a^{\dagger} \hat{L}_a \ket{-\psi_{n+1}^-}}{\braket{-\psi_{n}^-}{-\psi_{n+1}^-}} \, ,
\end{align}
and likewise for $H^{\pm}$. This gives the matrix element in \eqref{eq:rho_matrix_element} as
\begin{align}
    \bra{\psi_{n+1}} (1 + \Delta t \, \mathcal{L})  \Bigl( \dyad{\psi_{n}^+}{-\psi_n^-} \Bigr) \ket{-\psi_{n+1}^-} 
    & = \ee^{\bar{\psi}_{n+1}^+ \psi_{n}^+ + \bar{\psi}_{n}^{-} \psi_{n+1}^{-}} \left(1 + \Delta t \, \mathcal{L}[\bar{\psi}_{n+1}^{+}, \psi_{n}^+, \bar{\psi}_{n}^{-}, \psi_{n+1}^{-}] \right) \nonumber \\
    & \approx \ee^{\bar{\psi}_{n+1}^+ \psi_{n}^+ + \bar{\psi}_{n}^{-} \psi_{n+1}^{-}} \exp\left(\Delta t \, \mathcal{L}[\bar{\psi}_{n+1}^{+}, \psi_{n}^+, \bar{\psi}_{n}^{-}, \psi_{n+1}^{-}] \right) \, ,
\end{align}
where
\begin{align}
\mathcal{L}[\bar{\psi}^{+}, \psi^+, \bar{\psi}^{-}, \psi^{-}] = & \, 
- \ii H^+[\bar{\psi}^+, \psi] 
+ \ii H^-[-\bar{\psi}^{-}, -\psi^-] \nonumber \\
& + \sum_{a} \Bigl(
2 L_{a}^+[\bar{\psi}^{+}, \psi^+] \bar{L}_{a}^-[-\bar{\psi}^{-}, -\psi^-] 
- (\bar{L}_a L_{a})^+[\bar{\psi}^+, \psi^+] - (\bar{L}_a L_{a})^-[-\bar{\psi}^-, -\psi^-]  \Bigr) \, .
\end{align}
Taking the limit that $t_0 \rightarrow -\infty$ and $\mathcal{N} \rightarrow \infty$, we obtain
\begin{align}
Z = &\, \tr(\rho(T)) \nonumber \\
= &\, \int_{\psi_0^+ = - \psi_0^-} \!\! \mathcal{D} \bar{\psi}^+ \mathcal{D} \psi^+ \mathcal{D} \bar{\psi}^- \mathcal{D} \psi^- 
\exp\left[
\ii \int_0^T \dd{t} \Bigl(
\bar{\psi}^+ \ii \partial_t \psi^+ 
- \bar{\psi}^- \ii \partial_t \psi^- 
- \ii \mathcal{L}[\bar{\psi}^+, \psi^+, \bar{\psi}^-, \psi^-]\Bigr) 
\right] \nonumber \\
& \, \qquad \qquad \times \bra{\psi_{0}^+} \hat{\rho}(-\infty) \ket{-\psi_{0}^-} \, , \label{eq:Lindbladian_raw_form}
\end{align}
where the measure is defined as
\begin{equation}
    \int \mathcal{D} \bar{\psi}^+ \mathcal{D} \psi^+ \mathcal{D} \bar{\psi}^- \mathcal{D} \psi^- 
\equiv \lim_{\mathcal{N} \rightarrow \infty} \prod_{n=0}^{\mathcal{N}} \dd{\bar{\psi}_{n}^+} \dd{\psi_{n}^+} \dd{\bar{\psi}_{n}^-} \dd{\psi_{n}^-} \, .
\end{equation}

Let us then look at the Liouvillian functional \eqref{eq:Lindbladian_raw_form} and extract the minus signs from the terms $\ii H^-$, $2 L_a^+ \bar{L}_a^-$ and $(\bar{L}_a L_a)^-$: 
\begin{itemize}
    \item The $\ii H^-$ term: We assume that the Hamiltonian is parity even, such that 
    \[
        H^-[-\bar{\psi}^-, - \psi^-] = H^-[\bar{\psi}^-, \psi^-] \, .
    \] 
    \item The $(\bar{L}_a L_a)^-$ term: Since in $(\bar{L}_a L_a)^-$, $L_a$ and $\bar{L}_a$ appear in pairs, we have 
    \[ 
        (\bar{L}_a L_a)^-[- \bar{\psi}^-, -\psi^-] = (\bar{L}_a L_a)^-[\bar{\psi}^-, \psi^-] \, .
    \] 
    \item The $2 L_a^+ \bar{L}_a^-$ term: Denoting $\zeta = 1 \, (-1)$ if the jump operator is bosonic (fermionic), we have 
    \[ 
        \bar{L}_a^-[-\bar{\psi}^-, -\psi^-] = \zeta \bar{L}_a^-[\bar{\psi}^-, \psi^-] \, .
    \]
    Next, we perform a contour ordering which commutes $L_a^+$ and $\bar{L}_a^-$. The contour ordering then produces another $\zeta$ factor and therefore we get
    \[
        L_a^+[\bar{\psi}^+, \psi^+] \bar{L}_a^-[-\bar{\psi}^-, -\psi^-] = \bar{L}_a^-[\bar{\psi}^-, \psi^-] L_a^+[\bar{\psi}^+, \psi^+] \, .
    \]
\end{itemize}
To proceed, we can collectively write $z \equiv (t, \alpha)$ to denote the contour time, and hence $\dd{z} = \alpha \dd{t}$. The label $\alpha = \pm 1$ denotes which branch in the Schwinger-Keldysh contour the field is located in. Then for the contour integration of some function $f(z)$, we have
\begin{equation}
\int_{\mathcal{C}} f(z) \dd{z} = \sum_{\alpha= \pm } \int_{-\infty}^{\infty} f_{\alpha}(t) \, \alpha \dd{t} \, .
\end{equation}
Collecting these simplifications, and transforming back to the real Grassmann fields, the Schwinger-Keldysh path integral of the Lindbladian dynamics is given by
\begin{equation} \label{eq:Z_coherent_state}
    Z = \int \mathcal{D}a
    \exp\Biggl\{\ii \int_{\mathcal{C}} \dd{z} \frac{1}{2} \sum_{i} a_i(z) \ii \partial_z a_i(z) + \int \dd{t} \mathcal{L}[a^+(t), a^-(t)]\Biggr\} \, ,
\end{equation}
where the Liouvillian functional is
\begin{equation} \label{eq:Liouvillian_coherent_state}
    \mathcal{L} = - \ii H^+[a^+] + \ii H^-[a^-] + 2 \sum_{k} \bar{L}_k^-[a^-] L_k^+[a^+]  - \sum_{k} (\bar{L}_k L_k)^+[a^+] - \sum_{k} (\bar{L}_k L_k)^-[a^-] \, .
\end{equation}
Note that the recycling term is written in a contour ordered form, and there are no Bose/Fermi statistic factors. We can write the above form in a more compact way. Define the Lindblad kernel
\begin{equation} \label{eq:KL_def_appendix_A}
    \KL(z_1, z_2) \equiv \delta(t_1 - t_2) \KL_{\alpha_1 \alpha_2} \, ,
    \qquad
    \KL_{\alpha_1 \alpha_2} = \begin{pmatrix}
    1 & 0 \\
    2 & 1
    \end{pmatrix} \, , 
    \qquad
    \alpha_1, \alpha_2 = \pm \, ,
\end{equation}
we can write \eqref{eq:Z_coherent_state} as
\begin{equation}
    Z = \int \mathcal{D}a
    \exp\Biggl\{\ii \int_{\mathcal{C}} \dd{z} \biggl[ \frac{1}{2} \sum_{i} a_i(z) \ii \partial_z a_i(z) - H[a(z)] \biggr] - \int \dd{z_1} \dd{z_2} \KL(z_1, z_2) \bar{L}[a(z_1)] L[a(z_2)] \Biggr\} \, .
\end{equation}

\subsection{Path integral of the Lindblad SYK model}

The SYK Hamiltonian is given by
\begin{equation*}
H = \sum_{|I|=q} J_{I} \Gamma_{I} \, ,
\end{equation*}
where the random couplings $J_I \in \mathbb{R}$ are drawn from a Gaussian distribution with
\begin{equation*}
\langle J_I \rangle = 0 \, , \quad 
\langle J_I^2 \rangle = \frac{J^2 (q-1)!}{N^{q-1}} \, .
\end{equation*}
We consider the jump operators
\begin{equation*}
L_a = \sum_{\substack{|I| = k}} \ell_{I}^{a} \Gamma_{I} \, ,
\quad
a = 1, 2, \dots, M \, , \quad M = m N \, .
\end{equation*}
where $\ell_{I}^{a} \in \mathbb{C}$ are complex random numbers drawn from Gaussian distribution with \cite{sa2022,kulkarni2022}
\begin{equation*}
    \langle \ell_{I}^{a} \rangle = 0 \, , 
    \quad
    \langle |\ell_{I}^{a}|^2 \rangle = \frac{k! \gamma^2}{(2k)^{1/2} N^k} \, .
\end{equation*}
According to \eqref{eq:Liouvillian_coherent_state}, this gives the Liouvillian functional 
\begin{equation}
    \mathcal{L} = - \ii \sum_{|I| = q} J_{I} 
        \bigl(
        \Gamma_I^+ -  \Gamma_I^-
        \bigr) 
        + \sum_{a=1}^{M} \sum_{|I| = |J| = k} \bar{\ell}_I^a \ell_{J}^a \bigl(
        2 \Gamma_I^- \Gamma_J^+
        - \Gamma_I^+ \Gamma_J^+
        - \Gamma_I^- \Gamma_J^-
    \bigr) \, ,
    \quad
    \Gamma_I^{\alpha} \equiv \Gamma_I[\chi^{\alpha}(t)] \, .
\end{equation}
To proceed we define the Lindblad kernel on the Schwinger-Keldysh contour as in \eqref{eq:KL_def_appendix_A}
\begin{equation}
    \KL(z_1, z_2) \equiv \delta(t_1 - t_2) \KL_{\alpha_1 \alpha_2} \, ,
    \qquad
    \KL_{\alpha_1 \alpha_2} = \begin{pmatrix}
    1 & 0 \\
    2 & 1
    \end{pmatrix} \, , 
    \qquad
    \alpha_1, \alpha_2 = \pm \, ,
\end{equation}
and introduce a matrix functional $A_{IJ}[\chi]$
\begin{equation} \label{eq:A_IJ_chi}
A_{IJ}[\chi] \equiv - \int_{\mathcal{C}} \KL(z_1, z_2) \, \Gamma_I(z_1) \, \Gamma_J(z_2) \dd{z_1} \dd{z_2} \, .
\end{equation}
The path integral can be written as
\begin{equation}
    Z = \int \mathcal{D}\chi(z) \exp\Biggl\{
     \ii \int_{\mathcal{C}} \Biggl[ 
        \frac{\ii}{2} \sum_{j=1}^N \chi_j \partial_{z} \chi_j 
        - H 
    \Biggr] \dd{z}
    +
    \underbrace{\sum_{a=1}^{M} \sum_{|I|=|J|=k} \bar{\ell}_I^a \, A_{IJ}[\chi] \, \ell_{J}^{a}}_{\ii S_{\rm{L}}[\chi]} 
    \Biggr\}
\end{equation}
We then perform disorder averages over the random couplings. Defining the bi-local collective field 
\begin{equation} \label{eq:G_def_appendix}
    G(z, z') = - \frac{\ii}{N} \sum_{i=1}^{N} \chi_i(z) \chi_i(z') \, ,
\end{equation}
The disorder average over $J$ gives
\begin{equation}
    \left\langle \exp(- \ii \int_{\mathcal{C}} H \dd{z}) \right\rangle_{J}
     = \exp\left[
 - \frac{\ii^{q} N J^2}{2 q} \int_{\mathcal{C}} G(z_1, z_2)^{q} \dd{z_1} \dd{z_2}
\right] \, .
\end{equation}
The disorder average over $\ell_I$ gives
\begin{equation}
    \left\langle \exp(\ii S_{\rm{L}}[\chi]) \right\rangle_{\ell} 
    = \exp\bigl[- M \tr \log(1 - \langle |\ell_{I}^{a}|^2 \rangle A)\bigr] 
    = \exp \left[ M 
    \sum_{r=1}^{\infty} \frac{\langle |\ell_{I}^{a}|^2 \rangle^r}{r} \tr(A^r)
    \right] \, .
\end{equation}
Using definition of $A$ \eqref{eq:A_IJ_chi}, we have 
\begin{align}
\tr(A^r) & = (-1)^{r} \sum_{|I_1| = |I_2| = \dots = |I_r| = k} \int_{\mathcal{C}} \prod_{i=1}^{r} \Bigl[ \dd{z_i} \dd{z'_i} \, \KL(z_i, z'_i) \Gamma_{I_i}(z_i) \Gamma_{I_{i+1}}(z'_i) \Bigr] \\
& = (-1)^{r+k} \left(\frac{\ii^k N^k}{k!}\right)^{r} \int_{\mathcal{C}} \prod_{i=1}^r \Bigl[ \dd{z_i} \dd{z'_i} \KL(z_i, z'_i) [G(z'_i, z_{i+1})]^{k} \Bigr] \, ,
\end{align}
with $z_{r+1} = z_1$. Therefore,
\begin{equation} \label{eq:exp_iS_L_full_series}
    \left\langle \exp(\ii S_{\rm{L}}[\chi]) \right\rangle_{\ell} = 
    \exp \left[ N 
    \sum_{r=1}^{\infty} 
    (-1)^{r+k} \frac{m}{r} \biggl(\frac{\ii^k \gamma^2}{(2k)^{1/2}}\biggr)^{r} \int_{\mathcal{C}} \prod_{i=1}^r \Bigl( \dd{z_i} \dd{z'_i} \KL(z_i, z'_i) [G(z'_i, z_{i+1})]^{k} \Bigr)
    \right] \, .
\end{equation}
We then introduce the Lagrange multiplier $\Sigma(z, z')$ to impose the definition of the bi-local field \eqref{eq:G_def_appendix}, 
\begin{equation}
    1 = \int \mathcal{D} G \mathcal{D} \Sigma \exp\biggl\{
        - \frac{N}{2} \int_{\mathcal{C}} \Sigma(z_1, z_2) \biggl[
            G(z_1, z_2) + \frac{\ii}{N} \sum_{j=1}^{N} \chi_j(z_1) \chi_j(z_2)
        \biggr] \dd{z_1} \dd{z_2}
    \biggr\} \, ,
\end{equation}
and integrate out the Majorana fermions $\chi_j(z)$. This results in the large $N$ path integral on the Schwinger-Keldysh contour $\mathcal{C}$ after disorder average
\begin{equation}
Z = \int \mathcal{D} G \mathcal{D} \Sigma \ee^{\ii S[G, \Sigma]} \, ,
\end{equation}
where the action is
\begin{align} \label{eq:full_action_appendixB}
    \frac{2}{N} \ii S = &
    \Tr_{\mathcal{C}} \log (\ii \partial - \Sigma) 
    - \int_{\mathcal{C}} \Sigma(z, z') G(z, z') \dd{z} \dd{z'}
    - \frac{\ii^q J^2}{q} \int_{\mathcal{C}} [G(z, z')]^{q} \dd{z} \dd{z'} \nonumber \\
    & + 2 m \sum_{r=1}^{\infty} \frac{(-1)^{r+k}}{r} \biggl(\frac{\ii^k \gamma^2}{(2k)^{1/2}}\biggr)^{r} \int_{\mathcal{C}} \prod_{i=1}^{r} \dd{z_i} \dd{z_i'} \KL(z_i, z_i') [G(z_i', z_{i+1})]^{k} \, . 
\end{align}
If we keep only the first two terms $(r=1,2)$ of the action associated with the jump operators, as in \cite{sa2022,kulkarni2022}, the action is
\begin{align}
    \frac{2}{N} \ii S = &\,
    \Tr_{\mathcal{C}} \log (\ii \partial - \Sigma) 
    - \int_{\mathcal{C}} \Sigma(z, z') G(z, z') \dd{z} \dd{z'}
    - \frac{\ii^q J^2}{q} \int_{\mathcal{C}} [G(z, z')]^{q} \dd{z} \dd{z'} \nonumber \\
    &\, + \frac{m \gamma^4}{2 k} \int_{\mathcal{C}} \KL(z, z') \KL(w, w') [G(z', w)]^k [G(w', z)]^k \dd{z}\dd{z'}\dd{w}\dd{w'} \nonumber \\ 
    &\, - \frac{2 \ii^k m \gamma^2}{(2k)^{1/2}} \int_{\mathcal{C}} \KL(z, z') [G(z, z')]^{k} \dd{z} \dd{z'} \, .
\end{align}

\begin{figure}[!thb]
    \includegraphics[width=0.9\textwidth]{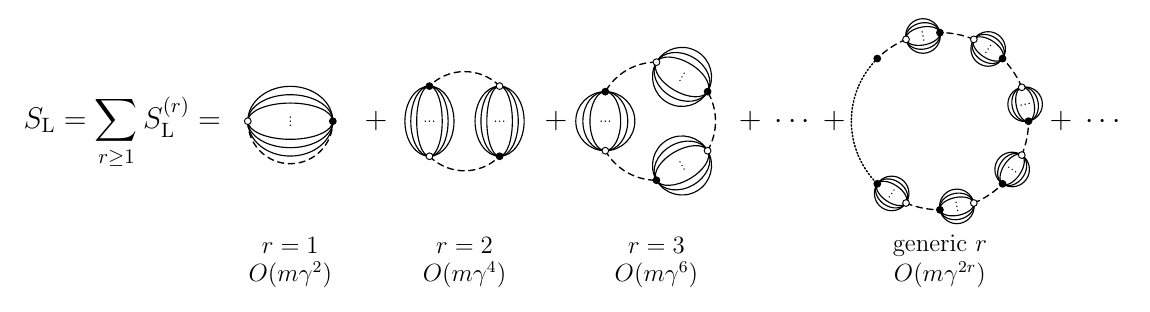}
    \caption{Expansion of the Lindblad action $S_{\rm{L}}$. Dashed lines denote the Lindblad kernel $\KL(z_1, z_2)$, and solid lines denote the bi-local field $G(z_1, z_2)$.}
\end{figure}

\subsection{Schwinger-Dyson equations}

\begin{figure}[!t]
    \includegraphics[width=0.32\textwidth]{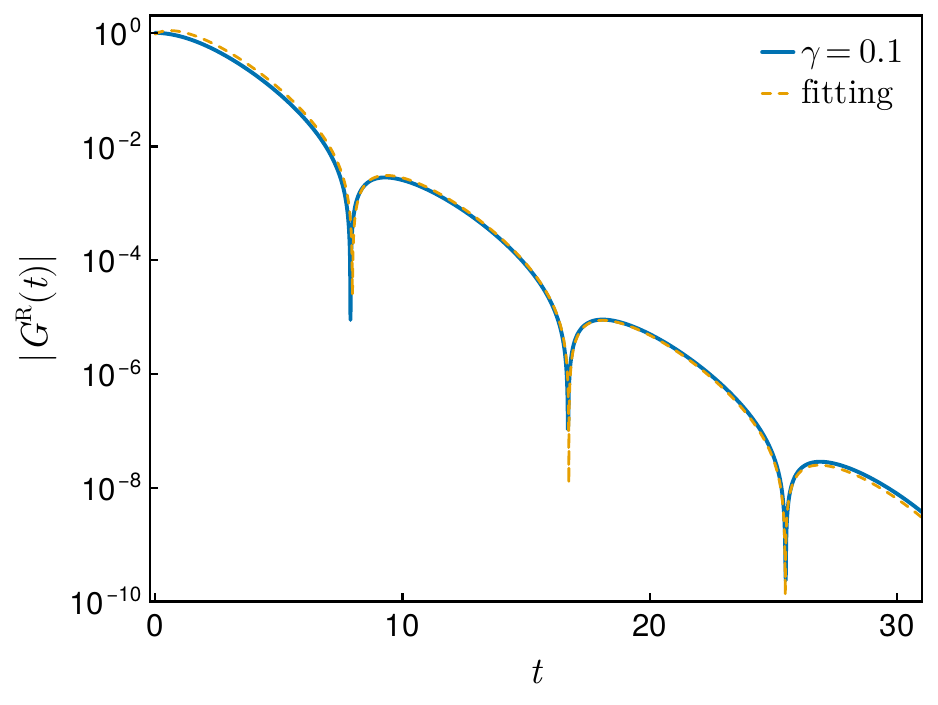}
    \includegraphics[width=0.32\textwidth]{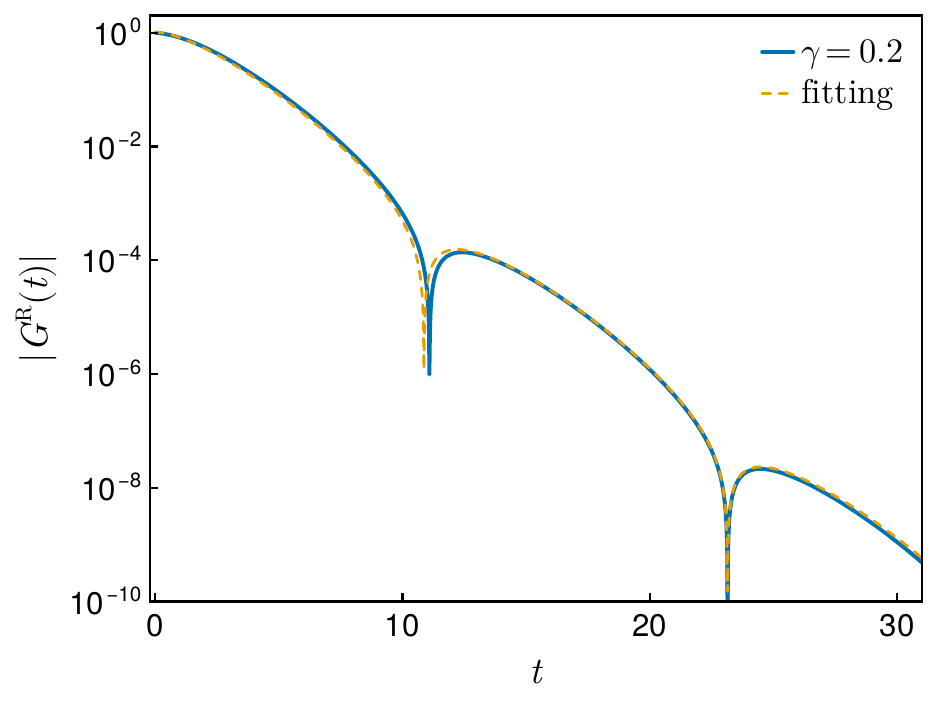}
    \includegraphics[width=0.32\textwidth]{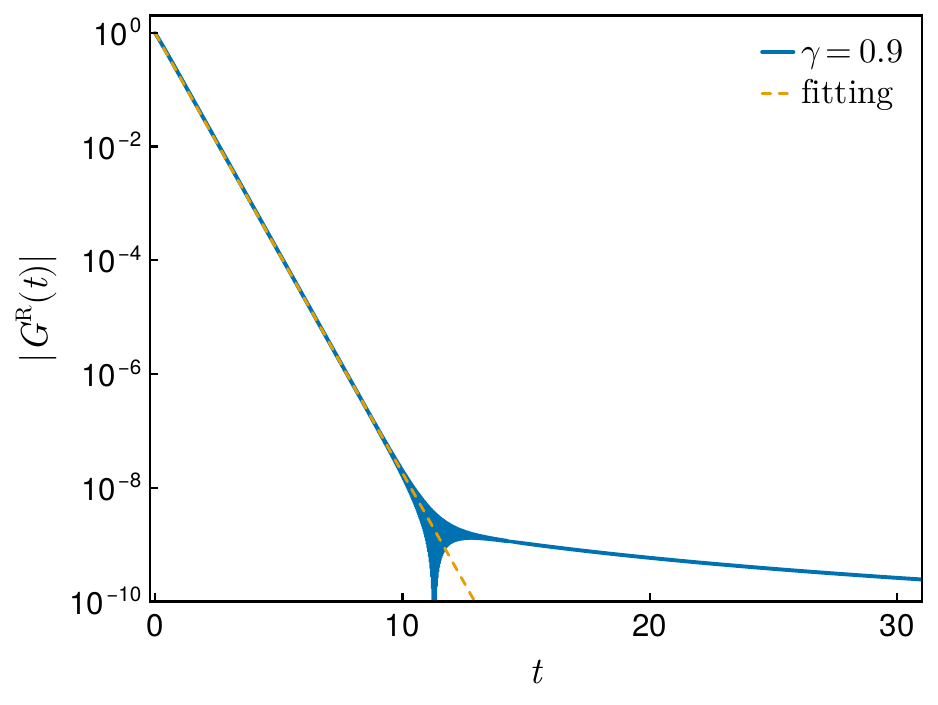}
    \caption{Retarded Green's functions at different dissipation strength for $q=4$, $m = 2$ and $k=2$.}
    \label{fig:GRt}
\end{figure}

\begin{figure}[!t]
    \includegraphics[width=0.45\textwidth]{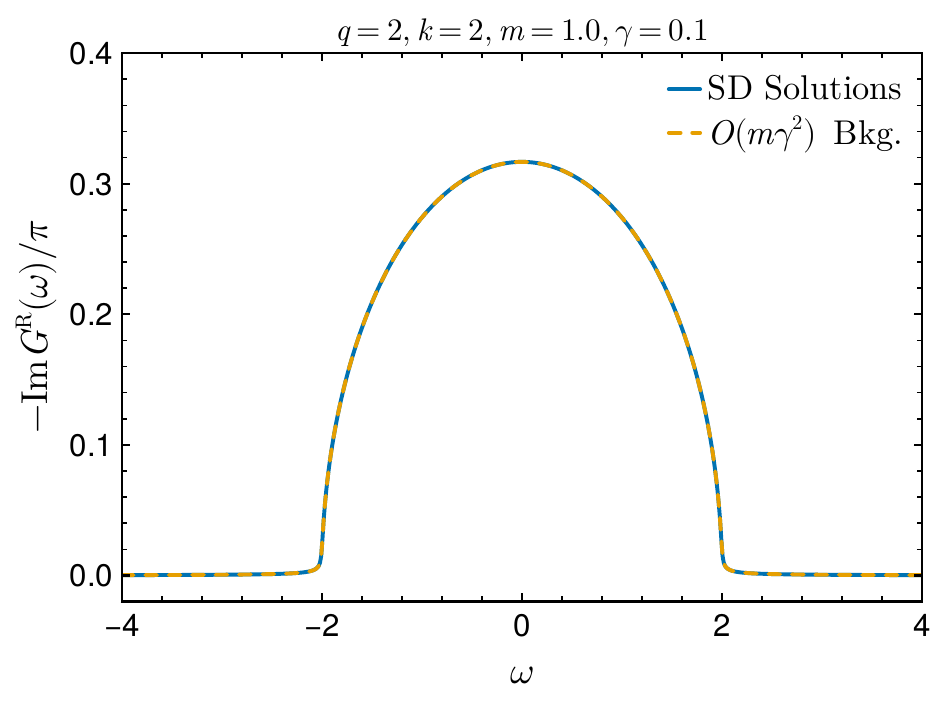}
    \includegraphics[width=0.45\textwidth]{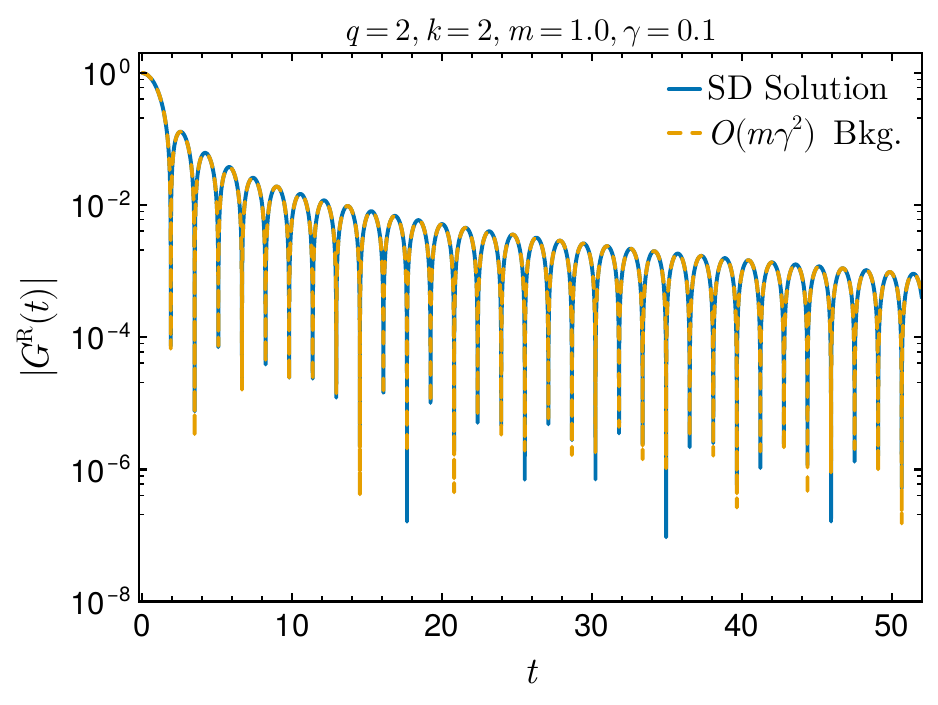}
    \caption{Comparisons between the numerical solutions of the SD equation with the analytic solution for $q=2$.}
\end{figure}

The large $N$ Schwinger-Dyson equations are obtained by varying the action with respect to $G$ and $\Sigma$.  Using the anti-symmetrized functional derivative 
\begin{equation}
    \fdv{G(w_1, w_2)}{G(z_1, z_2)} = \frac{1}{2} \bigl[\delta(w_1, z_1) \delta(w_2, z_2) - \delta(w_1, z_2) \delta(w_2, z_1)\bigr] \, ,
\end{equation}
the Schwinger-Dyson equations are
\begin{align}
    & (\ii \partial - \Sigma) \circ G = 1_{\mathcal{C}} \, , \\
    &  \Sigma(z_1, z_2) = - \ii^q J^2 [G(z_1, z_2)]^{q-1}+ m k G(z_1, z_2)^{k-1} \biggl[\mathcal{M}_1(z_1, z_2) + \sum_{r=2}^{\infty} \mathcal{M}_r(z_1, z_2) \biggr] \, ,
\end{align}
where $\circ$ denotes the convolution on the Schwinger-Keldysh contour, and
\begin{align}
    \mathcal{M}_1(z_1, z_2) = & \, - \frac{\ii^k \gamma^2}{(2k)^{1/2}} \bigl[\KL(z_1, z_2) + (-1)^k \KL(z_2, z_1)\bigr] \, ,\\
    \mathcal{M}_r(z_1, z_2) = & \, (-1)^{r+k} \biggl(\frac{\ii^k \gamma^2}{(2k)^{1/2}}\biggr)^{r} \int_{\mathcal{C}} \prod_{i=2}^{r-1} \dd{w_i} \dd{w_i'} \KL(w_i, w_i') [G(w_i', w_{i+1})]^{k} \nonumber \\
    & \times \int_{\mathcal{C}} \bigl[\KL(w_1, z_1) \KL(z_2, w_1') + (-1)^k \KL(w_1, z_2) \KL(z_1, w_1')\bigr] [G(w_1', w_2)]^k \dd{w_1} \dd{w_1'} \, ,
\end{align}
with $w_{r} \equiv w_1$. Note that both $G$ and $\Sigma$ are antisymmetric, $G(z_1, z_2) = - G(z_2, z_1)$ and likewise for $\Sigma$. Keeping only $r = 1, 2$, the self energy reads
\begin{align}
    \Sigma(z_1, z_2) = &\, 
    - \ii^{q} J^2 [G(z_1, z_2)]^{q-1}
    - \frac{\ii^k m k \gamma^2}{(2k)^{1/2}} \bigl[\KL(z_1, z_2) + (-1)^k \KL(z_2, z_1)\bigr] [G(z_1, z_2)]^{k-1} \nonumber \\
    &\, + \frac{m \gamma^4}{2} [G(z_1, z_2)]^{k-1} \int_{\mathcal{C}} \bigl[\KL(w_1, z_1) \KL(z_2, w_2) + (-1)^k\KL(w_1, z_2) \KL(z_1, w_2)\bigr] [G(w_2, w_1)]^{k} \dd{w_1} \dd{w_2} \, .
    \label{eq:Sigma_z1z2}
\end{align}
Projecting to $-+$ ($+-$) components gives the greater (lesser) self energy
\begin{align}
    \Sigma^>(t_1, t_2) & = 
    - \ii^q J^2 [G^>(t_1, t_2)]^{q-1}
    - m [G^>(t_1, t_2)]^{k-1} \biggl\{
        \ii^{k} (2k)^{\frac{1}{2}} \gamma^2 \delta(t_1 - t_2)
        + (-1)^{k} \gamma^4 [G^<(t_1, t_2)]^{k}
     \biggr\} \, , \\
    \Sigma^<(t_1, t_2) & = 
    - \ii^{q} J^2 [G^<(t_1, t_2)]^{q-1}
    - (-1)^{k} m [G^<(t_1, t_2)]^{k-1} \biggl\{
        \ii^{k} (2k)^{\frac{1}{2}} \gamma^2 \delta(t_1 - t_2)
        + \gamma^4 [G^>(t_1, t_2)]^{k}
     \biggr\} \, .
\end{align}
Note that this formula is consistent with Eq.~(S151) of \cite{sa2022} at $k=2$, but for general $k$ they do not agree. Note that both $G$ and $\Sigma$ are antisymmetric, $G(z_1, z_2) = - G(z_2, z_1)$ and similarly for $\Sigma$. Since the jump operators are Hermitian, the steady state is at infinite temperature with vanishing Keldysh Green's function $G^{\rm{K}} = G^> + G^< = 0$ due to the KMS condition $G^>(t_1, t_2) = - G^<(t_1, t_2)$. Together with the symmetry conditions $G^>(t_1, t_2) = - G^<(t_1, t_2)$ and $[G^>(t_1, t_2)]^{*} = G^<(t_1, t_2)$, we need only solve for the greater Green's function $G^>(t_1 - t_2)$ at the steady state, which is a purely imaginary even function. 

\subsection{The full action and Schwinger-Dyson equations}

\begin{figure}[!t]
    \includegraphics[width=0.4\textwidth]{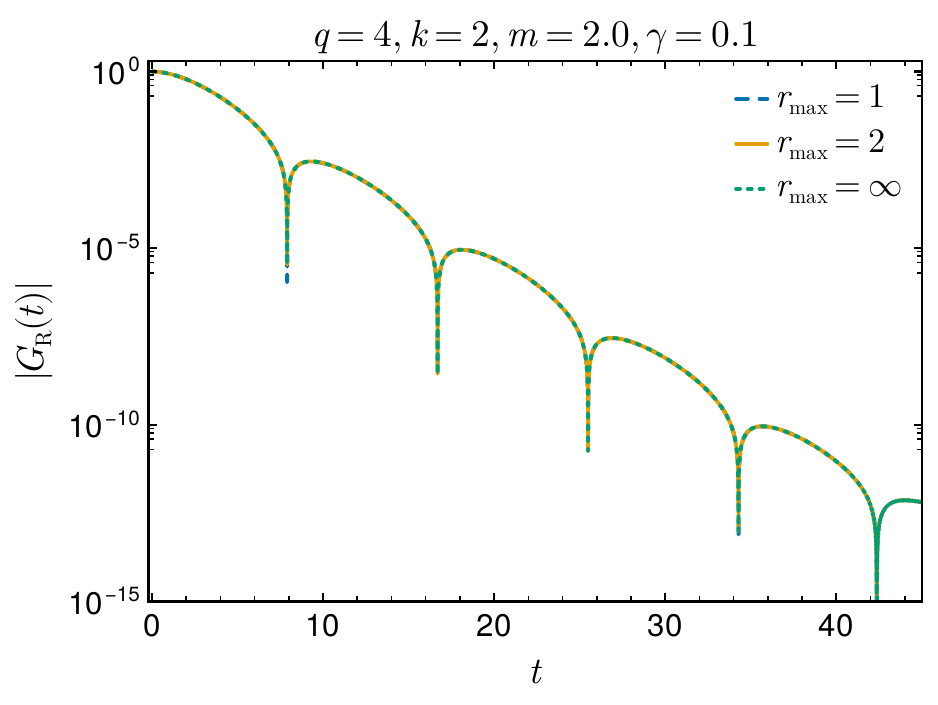}
    \hspace{1cm}
    \includegraphics[width=0.4\textwidth]{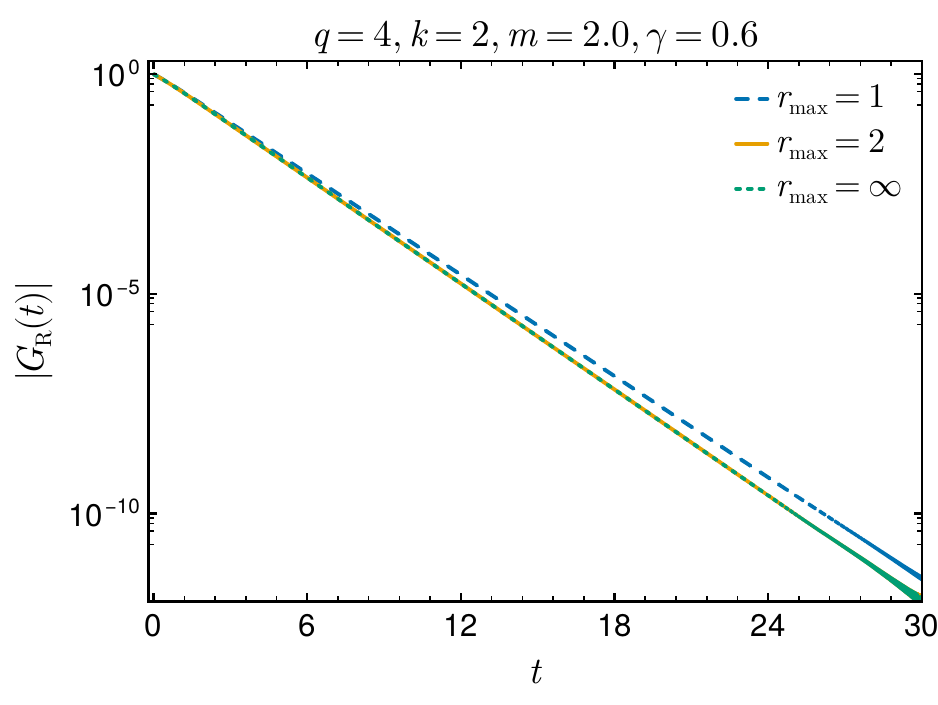}
    \caption{Comparison of retarded Green's functions for different truncations of the Lindblad action $S_{\rm{L}}$.}
    \label{fig:GRt_SL_truncation}
\end{figure}

Above we perform an expansion of $S_{\rm{L}}$ and truncate it at order $m \gamma^4$, as was down in \cite{sa2022,kulkarni2022}. It turns out that the infinite series expansion \eqref{eq:exp_iS_L_full_series} can be summed to give en exact result. Denoting
\begin{equation}
    B(z_1, z_2) \equiv \int_{\mathcal{C}} \KL(z_1, z) [G(z, z_2)]^{k} \dd{z} \, ,
\end{equation}
The Lindblad action \eqref{eq:exp_iS_L_full_series} becomes
\begin{align}
    \ii S_{\rm{L}} & = N \sum_{r=1}^{\infty} (-1)^{r + k} \frac{m}{r} \biggl(\frac{\ii^k \gamma^2}{(2k)^{1/2}}\biggr)^{r} \int_{\mathcal{C}} B(z_1, z_2) B(z_2, z_3) \dots B(z_r, z_1) \dd{z_1} \dots \dd{z_r} \nonumber \\
    & = N \sum_{r=1}^{\infty} (-1)^{r + k} \frac{m}{r} \biggl(\frac{\ii^k \gamma^2}{(2k)^{1/2}}\biggr)^{r} \Tr_{\mathcal{C}} (B^r) \nonumber \\
    & = - N (-1)^k m \Tr_{\mathcal{C}} \log(1 + \frac{\ii^k \gamma^2}{(2k)^{1/2}} B)
\end{align}
The full action therefore is
\begin{align}
    \frac{2}{N} \ii S = &\,
    \Tr_{\mathcal{C}} \log (\ii \partial - \Sigma) 
    - \int_{\mathcal{C}} \Sigma(z, z') G(z, z') \dd{z} \dd{z'}
    - \frac{\ii^q J^2}{q} \int_{\mathcal{C}} [G(z, z')]^{q} \dd{z} \dd{z'} \nonumber \\
    &\, - 2 (-1)^k m \Tr_{\mathcal{C}} \log(1 + \frac{\ii^k \gamma^2}{(2k)^{1/2}} B) \, .
\end{align}
The Schwinger-Dyson equations are given by
\begin{align}
    \label{eq:SD_full_1}
    & (\ii \partial - \Sigma) \circ G = 1_{\mathcal{C}} \, , \\
    \label{eq:SD_full_2}
    &  \Sigma(z_1, z_2) = - \ii^q J^2 [G(z_1, z_2)]^{q-1}
    - \ii^k (k/2)^{1/2} m \gamma^2 G(z_1, z_2)^{k-1} \biggl[\mathcal{H}(z_1, z_2) + (-1)^{k} \mathcal{H}(z_2, z_1) \biggr] \, , \\
    \label{eq:SD_full_3}
    & \biggl[1 + \frac{\ii^k \gamma^2}{(2k)^{1/2}} B \biggr] \circ \mathcal{H} = \KL \, .
\end{align}
This can be numerically solved in a similar manner with finite truncations. The numerical solutions of $G^{\rm{R}}(t)$ for different truncations are presented in Fig.~\ref{fig:GRt_SL_truncation}, which shows that the full action gives the same results with a truncation of $S_{\rm{L}}$ at order $m \gamma^4$. The higher order terms are therefore irrelevant. 

\section{Out-of-time-ordered correlation functions and Lyapunov exponents}

\subsection{Recursion relations of connected four-point functions}
\label{appendix_sub:Recursion_relations_of_connected_four-point_functions}

In this appendix, we present details on the derivation of the four-point functions on generic Keldysh contour~$\mathcal{C}$, which serves as the starting point for projections onto the OTOC kernels. We will follow the discussions and notations in \cite{Ferrari:2019ogc}. It is convenient to introduce a bi-local source~$\sJ(z_1, z_2)$, and the path integral after disorder average reads
\begin{equation}
    Z[\sJ] = \int \mathcal{D}G \mathcal{D} \Sigma \, \exp(\ii S[G, \Sigma, \sJ]) \, ,
\end{equation}
where the action is
\begin{equation}
    \frac{\ii}{N} S = \frac{1}{2} \mathrm{Tr}_{\mathcal{C}} \log(\ii \partial - \Sigma) - \frac{1}{2} \int_{\mathcal{C}} \Sigma(z_{1}, z_{2}) G(z_{1}, z_{2}) \dd z_{1} \dd z_{2} + s[G] - \ii \int_{\mathcal{C}} \sJ (z_{1}, z_{2}) G(z_{1}, z_{2}) \dd z_{1} \dd z_{2} \, ,
\end{equation}
with
\begin{align}
s[G] = 
- \frac{\ii^q J^2}{2 q} \int_{\mathcal{C}} [G(z_{1}, z_{2})]^{q} \dd z_{1} \dd z_{2} 
+ m \sum_{r=1}^{\infty} \frac{(-1)^{r+k}}{r} \biggl(\frac{\ii^k \gamma^2}{(2k)^{1/2}}\biggr)^{r} \int_{\mathcal{C}} \prod_{i=1}^{r} \dd{z_i} \dd{z_i'} \KL(z_i, z_i') [G(z_i', z_{i+1})]^{k}
\end{align}

the interactions of \eqref{eq:full_action_appendixB} including the SYK interaction and also the jump operator terms. In the Keldysh contour $z \in \mathcal{C}$, we can compute the $n$-point functions of the bi-local collective field $G$ (2$n$-point functions of the Majorana fermions) by taking functional derivatives of $Z[\sJ]$ with respect to $\sJ$ and then setting $\sJ$ to 0 afterwards,
\begin{equation} \label{eq:n_pt_function_source_formula}
\langle \operatorname{T}_{\mathcal{C}} G(z_{1}, z_{1}') G(z_{2}, z_{2}') \dots G(z_{n}, z_{n}') \rangle = 
\left. 
\frac{\ii^n}{N^n Z} \frac{\delta}{\delta \sJ(z_1, z_1')} \frac{\delta}{\delta \sJ(z_2, z_2')} \dots \frac{\delta}{\delta \sJ(z_n, z_n')} Z[\mathcal{\sJ}]
\right|_{\sJ=0} \, .
\end{equation}
The contour ordering $\operatorname{T}_{\mathcal{C}}$ is defined such that for a string of operators $\mathcal{O}(z_i)$, we have
\begin{equation}
    \operatorname{T}_{\mathcal{C}} \mathcal{O}(z_1) \mathcal{O}(z_2) \dots \mathcal{O}(z_n) = \mathcal{O}(z_1') \mathcal{O}(z_2') \dots \mathcal{O}(z_n') \, , 
\quad \Im z_1' < \Im z_2' < \dots < \Im z_n' \, .
\end{equation}
The large $N$ background with the source is given by the leading order Schwinger-Dyson equations

\begin{align} 
\label{eq:SD_equations_with_source_G}
0 & = \frac{\ii}{N} \frac{\delta S}{\delta G_0(z_1, z_2)} = - \frac{1}{2} \Sigma_0(z_1, z_2) + \frac{\delta s[G]}{\delta G_0(z_1, z_2)} - \sJ(z_1, z_2) \, , \\
\label{eq:SD_equations_with_source_Sigma}
0 & = \frac{\ii}{N} \frac{\delta S}{\delta \Sigma_0(z_1, z_2)} = - \frac{1}{2} (\ii \partial - \Sigma_0)^{-1}(z_1, z_2) - \frac{1}{2} G_0(z_1, z_2) \, .
\end{align}
Denoting $(G_0, \Sigma_0)$ as the saddle point solutions, this gives
\begin{align}
G_0(z_1, z_2) & = (\ii \partial - \Sigma_0)^{-1}(z_1, z_2) \, , \\
\Sigma_0(z_1, z_2) & = 2 \frac{\delta s[G]}{\delta G_0(z_1, z_2)} - 2 \ii \sJ(z_1, z_2) \, .
\end{align}
On the other hand, varying the action with respect to the bi-local source gives
\begin{equation} \label{eq:variation_action_bi_local_source}
    \frac{\delta \ii S_0}{\delta \sJ(z_1, z_2)} = - \ii N G_0(z_1, z_2) \, .
\end{equation}
The on-shell action therefore is an implicit functional of the bi-local source, $S_0  = S_0[G_0[\sJ], \Sigma_0[\sJ], \sJ]$. The partition function then is approximated by the saddle point solutions
\begin{equation}
    Z_0[\sJ] = \exp(\ii S_0[G_0[\sJ], \Sigma_0[\sJ], \sJ] + O(N^{-1})) \, .
\end{equation}
Using Eq.~\eqref{eq:n_pt_function_source_formula}, we can compute the four-point function of the bi-local collective field $G$ as

\begin{align}
\langle \operatorname{T}_{\mathcal{C}} G(z_1, z_2) G(z_3, z_4) \rangle 
& = \left. \frac{\ii^2}{N^2 Z_0} \frac{\delta}{\delta \sJ(z_1, z_2)} \frac{\delta}{\delta \sJ(z_3, z_4)} Z_0 \right|_{\sJ=0} \\
& = 
\left. \frac{\ii^2}{N^2} \frac{\delta \ii S_0}{\delta \sJ(z_1, z_2)} \frac{\delta \ii S_0}{\delta \sJ(z_3, z_4)} \right|_{\sJ=0} +
\left.
\frac{\ii^2}{N^2} \frac{\delta^2 \ii S_0}{\delta \sJ(z_1, z_2) \delta \sJ(z_3, z_4)}
\right|_{\sJ=0} \\
& = G_0(z_1, z_2) G_0(z_3, z_4) + \left. \frac{\ii}{N^2} \frac{\delta^2 S_0}{\delta \sJ(z_1, z_2) \delta \sJ(z_3, z_4)}
\right|_{\sJ=0}
\end{align}

Since we have the large $N$ factorization for single-trace operators (bi-local collective fields in SYK)
\begin{equation}
    \langle \operatorname{T}_{\mathcal{C}} G(z_1, z_2) G(z_3, z_4) \rangle = \langle G(z_1, z_2) \rangle \langle G(z_3, z_4) \rangle + \frac{1}{N} \mathcal{F}(z_1, z_2, z_3, z_4) \, ,
\end{equation}
where $\langle G \rangle = G_0$ is the saddle point solution of the Schwinger-Dyson equation, and the connected four-point function $\mathcal{F}$ accounts for the order $1/N$ effects. Comparing the two formulas, we see that the connected four-point function $\mathcal{F}$ is given by
\begin{align}
    \mathcal{F}(z_1, z_2, z_3, z_4) & = \left. \frac{\ii^2}{N} \frac{\delta^2 \ii S_0}{\delta \sJ(z_1, z_2) \delta \sJ(z_3, z_4)} \right|_{\sJ=0} \\[2pt]
    & = \ii \left. \frac{\delta G_0(z_1, z_2)}{\delta \sJ(z_3, z_4)} \right|_{\sJ = 0} \, .
\end{align}
To compute this functional derivative, we make use of the Schwinger-Dyson and use simplified notations $G_{ij} \equiv G(z_i, z_j)$ and repeated indices for integrations over the Keldysh contour, to obtain (The subsequent equations are meant to send the source $\sJ$ to 0 after the evaluation is done)
\begin{align*}
\mathcal{F}_{i j k l} & = \ii \left. \frac{\delta (G_0)_{ij}}{\delta \sJ_{kl}} \right|_{\sJ=0} \\
& = \ii \frac{\delta (\ii \partial - \Sigma_0)^{-1}_{ij}}{\sJ_{kl}}\\
& = - \ii [(\ii \partial - \Sigma_0)^{-1}]_{i m} \frac{\delta (\ii\partial - \Sigma_0)_{mn}}{\delta \sJ_{kl}} [(\ii \partial - \Sigma_0)^{-1}]_{n j} \\
& = \ii (G_0)_{i m} \frac{\delta \Sigma_{mn}}{\delta \sJ_{kl}} (G_0)_{n j} \\
& = \ii (G_0)_{i m} \left(2 \frac{\delta^2 s[G_0]}{\delta (G_0)_{mn} \delta \sJ_{kl}} - 2 \ii \frac{\delta \sJ_{mn}}{\delta \sJ_{kl}}\right) (G_0)_{n j}  \\
& = \ii (G_0)_{i m} \left(2 \frac{\delta^2 s[G_0]}{\delta (G_0)_{mn} \delta (G_0)_{rs}} \frac{\delta (G_0)_{rs}}{\delta \sJ_{kl}} - \ii \delta_{mk}  \delta_{nl} + \ii \delta_{ml} \delta_{nk} \right) (G_0)_{n j} \\
& = [(G_0)_{il} (G_0)_{jk} - (G_0)_{ik} (G_0)_{jl}] + 2 (G_0)_{i m} (G_0)_{n j} \frac{\delta^2 s[G_0]}{\delta (G_0)_{mn} \delta (G_0)_{rs}} \mathcal{F}_{r s k l} \, .
\end{align*}
Introducing the kernel $\mathcal{K}_{ijrs}$, 
\begin{equation}
    \mathcal{K}_{i j r s} \equiv 2 (G_0)_{i m} (G_0)_{n j} \frac{\delta^2 s[G_0]}{\delta (G_0)_{mn} \delta (G_0)_{rs}} \, ,
\end{equation}
and
\begin{equation}
    \mathcal{F}_{0, i j k l} \equiv (G_0)_{il} (G_0)_{jk} - (G_0)_{ik} (G_0)_{jl}
\end{equation}
we therefore find the recursion relation for $\mathcal{F}$
\begin{equation}
    \mathcal{F}_{i j k l} = \mathcal{F}_{0, i j k l} + \mathcal{K}_{i j r s} \mathcal{F}_{r s k l} \, .
\end{equation}
representing the summation of all ladder diagrams. In terms of the original notation, this reads
\begin{equation}
    \mathcal{F}(z_1, z_2; z_3, z_4) = \mathcal{F}_0(z_1, z_2; z_3, z_4) + \int_{\mathcal{C}} \mathcal{K}(z_1, z_2; z, z') \mathcal{F}(z, z'; z_3, z_4) \dd z \dd z' \, ,
\end{equation}
with the kernel given by
\begin{equation}
    \mathcal{K}(z_1, z_2; z_3, z_4) = 2 \int_{\mathcal{C}} G_0(z_1, z) G_0(z', z_2) \frac{\delta^2 s[G_0]}{\delta G_0(z, z') \delta G_0(z_3, z_4)} \dd z \dd z' \, .
\end{equation}
Using the Schwinger-Dyson equation without source, we can also write the kernel as
\begin{equation}
    \mathcal{K}(z_1, z_2; z_3, z_4) = \int_{\mathcal{C}} G_0(z_1, z) \fdv{\Sigma_0(z, z')}{G_0(z_3, z_4)} G_0(z' z_2) \dd{z} \dd{z'} \, .
\end{equation}

In our case, the kernel consists of the SYK interaction contribution and also the jump operators contribution:
\begin{equation}
    \mathcal{K} = \mathcal{K}_{J} + \mathcal{K}_{\rm{L}}^{(1)} + \mathcal{K}_{\rm{L}}^{(2)} + O(m \gamma^6) \, .
\end{equation}

Using the on-shell self energy Eq.~\eqref{eq:Sigma_z1z2}, we obtain
\begin{align}
    \label{eq:K_J_z}
    \mathcal{K}_{J} & = \ii^{q} (q-1) J^2 G_0(z_1, z_3) G_0(z_2, z_4) [G_0(z_3, z_4)]^{q-2} \, , \\
    \label{eq:K_L_r1}
    \mathcal{K}_{\rm{L}}^{(1)} & = \frac{\ii^k m k (k-1) \gamma^2}{(2k)^{1/2}} G_0(z_1, z_3) G_0(z_2, z_4) \bigl[\KL(z_3, z_4) + (-1)^k \KL(z_4, z_3)\bigr] [G_0(z_3, z_4)]^{k-2} \, , \\
    \label{eq:K_L_r2}
    \mathcal{K}_{\rm{L}}^{(2)} & = - \frac{m \gamma^4}{2} (k-1) G_0(z_1, z_3) G_0(z_2, z_4) [G_0(z_3, z_4)]^{k-2} \mathcal{P}(z_3, z_4) - \frac{m \gamma^4}{2} k \mathcal{Q}(z_1, z_2; z_3, z_4) [G_0(z_3, z_4)]^{k-1} \, ,
\end{align}
where
\begin{align}
    \label{eq:P_term_def}
    \mathcal{P}(z_3, z_4) & \equiv \int_{\mathcal{C}} \bigl[ \KL(w, z_3) \KL(z_4, w') + (-1)^{k} \KL(w, z_4) \KL(z_3, w') \bigr] [G(w', w)]^{k} \dd{w} \dd{w'} \, , \\
    \label{eq:Q_term_def}
    \mathcal{Q}(z_1, z_2; z_3, z_4) & \equiv \int_{\mathcal{C}} G_0(z_1, z) G_0(z_2, z') [G_0(z, z')]^{k-1} \bigl[ \KL(z_4, z) \KL(z', z_3) + (-1)^{k} \KL(z_4, z') \KL(z, z_3) \bigr] \dd{z} \dd{z'} \, .
\end{align}

\subsection{Projections onto the OTOC kernel}
\label{appendix_sub:OTOC_kernel_projections}

\begin{figure}[!thb]
\begin{center}
    \includegraphics[width=0.4\textwidth]{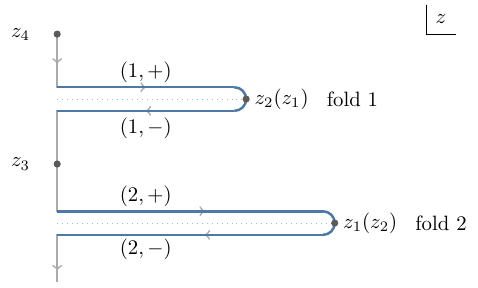}
    \caption{The OTOC contour.}
    \label{fig:OTOC_contour}
\end{center}
\end{figure}

In this appendix we derive in detail the projection of OTOC kernels from the general formulas \eqref{eq:K_J_z}, \eqref{eq:K_L_r1} and \eqref{eq:K_L_r2}. The OTOC contour consists of two time folds and is illustrated in Fig.~\ref{fig:OTOC_contour}. Consequently, each Majorana field $\chi$ should carry indices to indicate which fold (fold 1 or fold 2) and which branch (forward or backward) it is located on. For this purpose we introduce a compact index $z \equiv (t, A)$ with $A \equiv (a, \alpha)$, where $a \in \{1, 2\}$ labels the fold index, and $\alpha = \pm$ labels the branch index, so that we can write $\chi_i(z) \equiv \chi_i^A(t)$. Consequently, we may write the bi-local field as $G(z_1, z_2) \equiv G_{A_1 A_2}(t_1, t_2) \equiv G_{a_1 a_2}^{\alpha_1 \alpha_2}(t_1, t_2)$. With this notation the contour ordering is lexicographic
\begin{equation}
    (1, +) \prec (1, -) \prec (2, +) \prec (2, -) \, .
\end{equation}

We need also extend the previously defined one-time-fold Lindblad kernel \eqref{eq:K_lindblad_kernel_one_fold} to the OTOC contour. It is natural to generalize it as
\begin{equation} \label{eq:K_lindblad_kernel_OTOC_contour}
    \KL(z_1, z_2) = \delta(t_1 - t_2) \KL_{A_1 A_2} \, , 
    \quad 
    \KL_{A_1 A_2} = \begin{dcases}
        1 \, , & A_1 = A_2 \, , \\
        2 \, , & A_1 \succ A_2 \, , \\
        0 \, , & A_1 \prec A_2 \, .
    \end{dcases}
\end{equation}
Explicitly, 
\begin{equation}
    \KL_{AB} = \begin{pmatrix}
        1 & 0 & 0 & 0 \\
        2 & 1 & 0 & 0 \\
        2 & 2 & 1 & 0 \\
        2 & 2 & 2 & 1
    \end{pmatrix}_{AB} \, .
\end{equation}
Similarly, the Green's functions extend to
\begin{equation}
    G_{AB}(t_1, t_2) = G_{ab}^{\alpha \beta}(t_1, t_2) = \begin{dcases}
        G^{\alpha \beta}(t_1, t_2) \, , & a = b \, , \\
        G^{>}(t_1, t_2) \, , & a \succ b \, , \\
        G^{<}(t_1, t_2) \, , & a \prec b \, .
    \end{dcases}
\end{equation}
In the finite temperature case, $a \succ b$ and $a \prec b$ give the Wightman functions. However, in our case the steady state is infinite temperature state with $\beta = 0$, and all the imaginary times are sent to $0$ with their relative order fixed. As a result, the Wightman functions reduce to greater or lesser Green's functions depending on the ordering \cite{Garcia-Garcia:2024tbd,liu2026}.

There are two possible cases for the connected OTOCs, depending on which folds $z_1 \equiv (t_1, A_1)$ and $z_2 \equiv (t_2, A_2)$ live on. For this purpose we denote two OTOCs
\begin{align}
    & F_1(t_1, t_2) \, , \quad \text{with} \quad z_1 \in \text{fold 2} \, , \quad z_2 \in \text{fold 1} \, , \quad \Leftrightarrow \quad A_1 = (2, 0) \, , \quad A_2 = (1, 0) \, , \\
    & F_2(t_1, t_2) \, , \quad \text{with} \quad z_1 \in \text{fold 1} \, , \quad z_2 \in \text{fold 2} \, , \quad \Leftrightarrow \quad A_1 = (1, 0) \, , \quad A_2 = (2, 0) \, .
\end{align}
Here we use $0$ to denote that the fields are located on the tip of the time folds that separates the forward and backward branches. The connected OTOCs obey the recursion relations
\begin{equation}
    \label{eq:connected_OTOC_recursion_relation}
    F_i(t_1, t_2) = F_{i, 0}(t_1, t_2) +  \sum_{j=1}^{2} \int_{-\infty}^{\infty}\mathcal{K}_{i j}(t_1, t_2, t_3, t_4) F_j(t_3, t_4) \dd{t_3} \dd{t_4} \, .
\end{equation}
The OTOC kernels $\mathcal{K}_{ij}$ are obtained from projections of $\mathcal{K}(z_1, z_2; z_3, z_4)$ (\eqref{eq:K_J_z}, \eqref{eq:K_L_r1} and \eqref{eq:K_L_r2}) and summing over the branch indices of $z_3$ and $z_4$. They decomposes into 4 pieces: 
\begin{equation}
    \mathcal{K}_{i j} = \mathcal{K}_{J, i j} + \mathcal{K}_{{\rm{L}}, i j}^{(1)} + \mathcal{K}_{{\rm{L}}, \mathcal{P}, i j}^{(2)} + \mathcal{K}_{{\rm{L}}, \mathcal{Q}, i j}^{(2)} \, .
\end{equation}
For projection, let us denote
\begin{equation}
    \mathcal{K}(z_1, z_2; z_3, z_4) \equiv \mathcal{K}_{(a_1 a_2)(a_3 a_4)}^{(\alpha_1 \alpha_2)(\alpha_3 \alpha_4)}(t_1, t_2; t_3, t_4) \, ,
\end{equation}
the OTOC kernels in 
 is given by
\begin{align}
    \mathcal{K}_{11}(t_1, t_2; t_3, t_4) & = \theta(t_{13}) \theta(t_{24}) \sum_{\alpha_3, \alpha_4 = \pm} \alpha_3 \alpha_4 \mathcal{K}_{(21)(21)}^{(00)(\alpha_3 \alpha_4)}(t_1, t_2; t_3, t_4) \, , \\ 
    \mathcal{K}_{12}(t_1, t_2; t_3, t_4) & = \theta(t_{14}) \theta(t_{23}) \sum_{\alpha_3, \alpha_4 = \pm} \alpha_3 \alpha_4 \mathcal{K}_{(21)(12)}^{(00)(\alpha_3 \alpha_4)}(t_1, t_2; t_3, t_4) \, , \\ 
    \mathcal{K}_{21}(t_1, t_2; t_3, t_4) & = \theta(t_{14}) \theta(t_{23}) \sum_{\alpha_3, \alpha_4 = \pm} \alpha_3 \alpha_4 \mathcal{K}_{(12)(21)}^{(00)(\alpha_3 \alpha_4)}(t_1, t_2; t_3, t_4) \, , \\ 
    \mathcal{K}_{22}(t_1, t_2; t_3, t_4) & = \theta(t_{13}) \theta(t_{24}) \sum_{\alpha_3, \alpha_4 = \pm} \alpha_3 \alpha_4 \mathcal{K}_{(12)(12)}^{(00)(\alpha_3 \alpha_4)}(t_1, t_2; t_3, t_4) \, .
\end{align}
Since we are interested in the OTOC kernel, $z_3$ and $z_4$ are assigned to be located on different time folds. If $z_3$ and $z_4$ live on the same fold, we have the ordinary time ordered kernels which do not contribute to the OTOC ones. The pre-factors involving $\theta$-functions implement the causal restriction.

To warm up, let us take the kernel \eqref{eq:K_J_z} that corresponds to the SYK interaction. Using relation 
\begin{equation}
    \theta(t_{13}) \sum_{\alpha_3} \alpha_3 G^{0\alpha_3}(t_{13}) = \theta(t_{13}) [G^>(t_{13}) - G^<(t_{13})] = G^{\rm{R}}(t_{13}) \,
\end{equation}
we obtain
\begin{align}
    \mathcal{K}_{J, 11}(t_1, t_2; t_3, t_4) & = \ii^{q} (q-1) J^2 G^{\rm{R}}_0(t_{13}) G^{\rm{R}}_0(t_{24}) [G^>_0(t_{34})]^{q-2} \, , \\
    \mathcal{K}_{J, 22}(t_1, t_2; t_3, t_4) & = \ii^{q} (q-1) J^2 G^{\rm{R}}_0(t_{13}) G^{\rm{R}}_0(t_{24}) [G^<_0(t_{34})]^{q-2} \, , \\
    \mathcal{K}_{J, 12}(t_1, t_2; t_3, t_4) & = \mathcal{K}_{J, 21}(t_1, t_2; t_3, t_4) = 0 \, .
\end{align}
Similarly, for the kernel \eqref{eq:K_L_r1} we have
\begin{align}
    \mathcal{K}_{\rm{L}, 11}^{(1)}(t_1, t_2; t_3, t_4) & = \ii^{k} (2k)^{1/2} (k-1) m \gamma^2 G^{\rm{R}}_0(t_{13}) G^{\rm{R}}_0(t_{24}) [G^>_0(t_{34})]^{k-2} \delta(t_3 - t_4) \, , \\
    \mathcal{K}_{\rm{L}, 22}^{(1)}(t_1, t_2; t_3, t_4) & = (-1)^{k}\ii^{k} (2k)^{1/2} (k-1) m \gamma^2 G^{\rm{R}}_0(t_{13}) G^{\rm{R}}_0(t_{24}) [G^<_0(t_{34})]^{k-2} \delta(t_3 - t_4) \, , \\
    \mathcal{K}_{\rm{L}, 12}^{(1)}(t_1, t_2; t_3, t_4) & = \mathcal{K}_{\rm{L}, 21}^{(1)}(t_1, t_2; t_3, t_4) = 0 \, ,
\end{align}
where we used the extended Lindblad kernel \eqref{eq:K_lindblad_kernel_OTOC_contour}.

We then consider the OTOC projection for $\mathcal{K}_{\rm{L}}^{(2)}$ \eqref{eq:K_L_r2}. The first piece involves $\mathcal{P}(z_3, z_4)$ and can be easily worked out akin to the calculations in the previous two cases:
\begin{align}
    \mathcal{K}_{\rm{L}, \mathcal{P}, 11}^{(2)}(t_1, t_2; t_3, t_4) & = (-1)^{k} (k-1) m \gamma^4 G^{\rm{R}}_0(t_{13}) G^{\rm{R}}_0(t_{24}) [G^>_0(t_{34})]^{k-2} [G^<_0(t_{34})]^{k} \, , \\
    \mathcal{K}_{\rm{L}, \mathcal{P}, 22}^{(2)}(t_1, t_2; t_3, t_4) & = (-1)^{k} (k-1) m \gamma^4 G^{\rm{R}}_0(t_{13}) G^{\rm{R}}_0(t_{24}) [G^<_0(t_{34})]^{k-2} [G^>_0(t_{34})]^{k} \, , \\
    \mathcal{K}_{\rm{L}, \mathcal{P}, 12}^{(2)}(t_1, t_2; t_3, t_4) & = \mathcal{K}_{\rm{L}, \mathcal{P}, 21}^{(2)}(t_1, t_2; t_3, t_4) = 0 \, .
\end{align}
The second piece involves $\mathcal{Q}$ and gives
\begin{align}
    \mathcal{K}_{{\rm{L}}, \mathcal{Q}, 11}^{(2)}(t_1, t_2; t_3, t_4) & = \frac{(-1)^{k}}{2} k m \gamma^4 G^{\rm{R}}_0(t_{13}) G^{\rm{R}}_0(t_{24}) [G^>_0(t_{34})]^{2k-2} \, , \\
    \mathcal{K}_{{\rm{L}}, \mathcal{Q}, 12}^{(2)}(t_1, t_2; t_3, t_4) & = - \frac{(-1)^k}{2} k m \gamma^4 G^{\rm{R}}_0(t_{14}) G^{\rm{R}}_0(t_{23}) [G^<_0(t_{34})]^{2k-2} \, , \\
    \mathcal{K}_{{\rm{L}}, \mathcal{Q}, 21}^{(2)}(t_1, t_2; t_3, t_4) & = - \frac{(-1)^k}{2} k m \gamma^4 G^{\rm{R}}_0(t_{14}) G^{\rm{R}}_0(t_{23}) [G^>_0(t_{34})]^{2k-2} \, , \\
    \mathcal{K}_{{\rm{L}}, \mathcal{Q}, 22}^{(2)}(t_1, t_2; t_3, t_4) & = \frac{(-1)^k}{2} k m \gamma^4 G^{\rm{R}}_0(t_{13}) G^{\rm{R}}_0(t_{24}) [G^<_0(t_{34})]^{2k-2} \, .
\end{align}
In deriving these projections we use the relation
\begin{equation}
    [G^{\rm{T}}(t_1, t_2)]^{k} + [G^{\bar{\rm{T}}}(t_1, t_2)]^{k} = [G^>(t_1, t_2)]^{k} + [G^{<}(t_1, t_2)]^{k} \, .
\end{equation}
Computing the summations manually are cumbersome and we can do it simply in \texttt{Mathematica}.

In summary, denoting $G^{\rm{R}}_{i j} \equiv G^{\rm{R}}_0(t_{ij})$ and $G^{\gtrless}_{i j} \equiv G^{\gtrless}_0(t_{ij})$, we have the full OTOC kernels
\begin{align}
    \mathcal{K}_{11}(t_1, t_2; t_3, t_4) & = G^{\rm{R}}_{13} G^{\rm{R}}_{24} \begin{aligned}[t]
        \biggl\{ 
            & \ii^{q} (q-1) J^2 \bigl[G^>_{34}\bigr]^{q-2}
            + \ii^{k} (2k)^{\frac{1}{2}} (k-1) m \gamma^2 \bigl[G^>_{34}\bigr]^{k-2} \delta(t_{34}) 
            \\
            & + (-1)^{k}(k-1) m \gamma^4 \bigl[G^>_{34}\bigr]^{k-2} \bigl[G^<_{34}\bigr]^{k}
            + \frac{(-1)^k}{2} k m \gamma^4 \bigl[G^>_{34}\bigr]^{2k-2}
        \biggr\} \, ,
    \end{aligned} 
    \\
    \mathcal{K}_{22}(t_1, t_2; t_3, t_4) & = G^{\rm{R}}_{13} G^{\rm{R}}_{24} \begin{aligned}[t]
        \biggl\{ 
            & \ii^{q} (q-1) J^2 \bigl[G^<_{34}\bigr]^{q-2}
            + (-1)^{k} \ii^{k} (2k)^{\frac{1}{2}} (k-1) m \gamma^2 [G^<_{34}]^{k-2} \delta(t_{34})
            \\
            & + (-1)^k (k-1) m \gamma^4 \bigl[G^<_{34}\bigr]^{k-2} \bigl[G^>_{34}\bigr]^{k}
            + \frac{(-1)^k}{2} k m \gamma^4 \bigl[G^<_{34}\bigr]^{2k-2}
        \biggr\} \, ,
    \end{aligned} 
    \\
    \mathcal{K}_{12}(t_1, t_2; t_3, t_4) & = - \frac{(-1)^k}{2} k m \gamma^4 G^{\rm{R}}_{14} G^{\rm{R}}_{23} \bigl[G^<_{34}\bigr]^{2k-2} \, ,  
    \\
    \mathcal{K}_{21}(t_1, t_2; t_3, t_4) & = - \frac{(-1)^k}{2} k m \gamma^4 G^{\rm{R}}_{14} G^{\rm{R}}_{23} \bigl[G^>_{34}\bigr]^{2k-2} \, .
\end{align}

\subsection{Exponential growth ansatz and Lyapunov exponents}

To extract the Lyapunov exponents, we consider the long time limit of the connected OTOCs so that $F_{i,0}$ can be neglected, giving the recursion relations
\begin{equation}
    F_i(t_1, t_2) = \sum_{j=1}^{2} \int_{-\infty}^{\infty} \mathcal{K}_{i j}(t_1, t_2; t_3, t_4) F_j(t_3, t_4) \dd{t_3} \dd{t_4} \, .
\end{equation}
Taking an exponential growth ansatz
\begin{equation}
    F_i(t_1, t_2) = \ee^{\lambda T} f_i(t) \, , \quad
    T \equiv \frac{t_1 +t_2}{2} \, , \quad
    t \equiv t_1 - t_2 \, ,
\end{equation}
the recursion relations reduce to
\begin{equation}
    f_i(t) = \sum_{j=1}^{2} \int_{-\infty}^{\infty} M_{i j}(t, t' ; \lambda) f_j(t') \dd{t'} \, .
\end{equation}
Define
\begin{equation}
    A(t; \lambda) \equiv \int_{-\infty}^{\infty} G_0^{\rm{R}}\biggl(u + \frac{1}{2} t\biggr) G_0^{\rm{R}}\biggl(u - \frac{1}{2}t\biggr) \ee^{-\lambda u} \dd{u} \, ,
\end{equation}
the reduced kernel reads
\begin{align}
    M_{i j}(t, t' ; \lambda) = 
    \begin{dcases}
        \mathcal{R}_{i i}(t') A(t - t' ; \lambda) \, , & i = j \, , \\
        \mathcal{R}_{i j}(t') A(t + t' ; \lambda) \, , & i \neq j \, . \\
    \end{dcases}
\end{align}
where the rung factors $\mathcal{R}_{ij}$ are 
\begin{align}
    \mathcal{R}_{11}(t) & = \begin{aligned}[t]
        \biggl\{ 
            & \ii^{q} (q-1) J^2 \bigl[G^>(t)\bigr]^{q-2}
            + \ii^{k} (2k)^{\frac{1}{2}} (k-1) m \gamma^2 \bigl[G^>(t)\bigr]^{k-2} \delta(t) 
            \\
            & + (-1)^{k}(k-1) m \gamma^4 \bigl[G^>(t)\bigr]^{k-2} \bigl[G^<(t)\bigr]^{k}
            + \frac{(-1)^k}{2} k m \gamma^4 \bigl[G^>(t)\bigr]^{2k-2}
        \biggr\} \, ,
    \end{aligned} 
    \\
    \mathcal{R}_{22}(t) & = \begin{aligned}[t]
        \biggl\{ 
            & \ii^{q} (q-1) J^2 \bigl[G^<(t)\bigr]^{q-2}
            + (-1)^{k} \ii^{k} (2k)^{\frac{1}{2}} (k-1) m \gamma^2 [G^<(t)]^{k-2} \delta(t)
            \\
            & + (-1)^k (k-1) m \gamma^4 \bigl[G^<(t)\bigr]^{k-2} \bigl[G^>(t)\bigr]^{k}
            + \frac{(-1)^k}{2} k m \gamma^4 \bigl[G^<(t)\bigr]^{2k-2}
        \biggr\} \, ,
    \end{aligned} 
    \\
    \mathcal{R}_{12}(t) & = - \frac{(-1)^k}{2} k m \gamma^4 \bigl[G^<(t)\bigr]^{2k-2} \, ,  
    \\
    \mathcal{R}_{21}(t) & = - \frac{(-1)^k}{2} k m \gamma^4 \bigl[G^>(t)\bigr]^{2k-2} \, .
\end{align}
Here the cross rung factors $\mathcal{R}_{12}$ and $\mathcal{R}_{21}$ are due to the projections of the $\mathcal{Q}$ term \eqref{eq:Q_term_def}. The Lyapunov exponent $\lambda_{\rm{L}}$ is determined by the condition that the reduced kernel matrix $M$ has its maximal eigenvalue to be 1.

\end{document}